\documentclass[useAMS,usenatbib]{mn2e}
\pdfoutput=1

\usepackage{graphicx,color}
\usepackage{bm}
\usepackage{natbib}
\usepackage{amsmath}
\usepackage{xspace}
\usepackage{hyperref}
\usepackage[T1]{fontenc}
\usepackage{aecompl}

\newcommand{\science}{{Science\ }}
\newcommand{\nat}{{Nature\ }}
\newcommand{\aap}{{Astron.\ Astrophys.\ }}
\newcommand{\aj}{{Astron.\ J.\  }}
\newcommand{\apj}{{Astrophys.\ J.\  }}
\newcommand{\apjl}{{Astrophys.\ J.\ Lett.\  }}
\newcommand{\apjs}{{Astrophys.\ J.\ Suppl.\  }}
\newcommand{\mnras}{{Mon.\ Not.\ R.\ Astron.\ Soc.\ }}

\newcommand{\physrep}{{Phys.\ Rep.}}

\newcommand{\prd}{{Phys.\ Rev.\ D\ }}
\newcommand{\jcap}{{J. Cosmol. Astropart. Phys.}}

\def\sigv{\langle \sigma v\rangle}
\def\la{\langle}

\def\Ms{\, h^{-1} \, {\rm M}_{\odot}}
\def\Mpc{\, h^{-1} \, {\rm Mpc}}

\def\kMpc{\, h \, {\rm Mpc}^{-1}}

\def\kmax{k_{\rm max}}
\def\knl{k_{\rm NL}}
\def\neff{n_{s, {\rm eff}}}

\newcommand{\be}{\begin{equation}}
\newcommand{\ee}{\end{equation}}
\newcommand{\bdm}{\begin{displaymath}}
\newcommand{\edm}{\end{displaymath}}
\newcommand{\bea}{\begin{eqnarray}}
\newcommand{\eea}{\end{eqnarray}}
\newcommand{\bt}{\begin{tabular}}
\newcommand{\et}{\end{tabular}}
\newcommand{\de}{{d\;\!}}

\newcommand{\gsi}{\,\raisebox{-0.13cm}{$\stackrel{\textstyle>}{\textstyle\sim}$}\,}
\newcommand{\halofit}{{\texttt{halofit}}}

\title[Extragalactic gamma-ray signal from dark matter annihilation: an appraisal]{Extragalactic gamma-ray signal from dark matter annihilation: an appraisal}

\author[E.~Sefusatti, G.~Zaharijas, P.~D.~Serpico, D.~Theurel, and M.~Gustafsson]
  {E. Sefusatti$^{1,2}$\thanks{E-mail: emiliano.sefusatti@brera.inaf.it (ES)}, 
  G. Zaharijas$^{2,3}$\thanks{E-mail: gzaharij@ictp.it (GZ)},
  P. D. Serpico$^{4}$\thanks{E-mail: serpico@lapth.cnrs.fr (PDS)}, 
  D. Theurel$^{5}$\thanks{E-mail: theurel@mit.edu (DT)}, 
  \newauthor 
  and M. Gustafsson$^{6}$\thanks{E-mail: mgustafs@ulb.ac.be (MG)} \\
$^{1}$INAF - Osservatorio Astronomico di Brera, via E. Bianchi 46, I-23807 Merate (LC) -- Italy\\
$^{2}$The Abdus Salam International Center for Theoretical Physics, strada costiera 11, I-34151 Trieste -- Italy\\
$^{3}$INFN, Sezione di Trieste, via Valerio 2, I-34127 Trieste -- Italy\\
$^{4}$LAPTh, UMR 5108, 9 chemin de Bellevue - BP 110, 74941 Annecy-Le-Vieux, France\\
$^{5}$Center for Theoretical Physics, Massachusetts Institute of Technology, 77 Massachusetts Ave, 6-304, Cambridge, MA 02139 -- USA\\
$^{6}$Service de Physique Th\'eorique, Universit\'e Libre de Bruxelles, B-1050 Bruxelles -- Belgium}

\pagerange{\pageref{firstpage}--\pageref{lastpage}} \pubyear{????}
\date{\today}

\begin{document}
\maketitle
\label{firstpage}

\begin{abstract}
We re-evaluate the extragalactic gamma-ray flux prediction from dark matter annihilation in the approach of integrating over the nonlinear matter power spectrum, extrapolated to the free-streaming scale. We provide an estimate of the uncertainty based entirely on available N-body simulation results and minimal theoretical assumptions. We illustrate how an improvement in the simulation resolution, exemplified by the comparison between the Millennium and Millennium II simulations, affects our estimate of the flux uncertainty and we provide a Òbest guessÓ value for the flux multiplier, based on the assumption of stable clustering for the dark matter perturbations described as a collision-less fluid. We achieve results comparable to traditional Halo Model calculations, but with a much simpler procedure and a more general approach, as it relies only on one, {\em directly measurable} quantity. In addition we discuss the extension of our calculation to include baryonic effects as modeled in hydrodynamical cosmological simulations and other possible sources of uncertainty that would in turn affect indirect dark matter signals. Upper limit on the integrated power spectrum from supernovae lensing magnification are also derived and compared with theoretical expectations.
\end{abstract}

\section{Introduction}

Despite the great successes of the concordance cosmology model, the interpretation of its fit parameters (dark energy, dark matter, inflationary parameters, the very baryon fraction)
in terms of microphysics remains mysterious. For example, while a precise determination of the dark matter (DM) fraction of the energy budget of the universe has been achieved, we have not
identified its particle physics nature. Among the plethora of models proposed until now, weakly interacting massive particles (WIMPs) have the peculiarity of offering many potential signals
at colliders, direct detection experiments underground, or via astrophysical messengers coming from their residual annihilations. In particular, these annihilation processes should proceed not only in our Galactic Halo, but in the universe as a whole. Depending on the level of DM cross section and on the distribution of DM in cosmic structures, this process might be observed as a diffuse background. 

In this article we focus  on the computation of this Extragalactic DM annihilation Flux (EDMF), as a followup of our previous work in \citet{SerpicoEtal2012}. Although this signal has been usually calculated in the Halo Model (HM) framework \citep[see][]{UllioEtal2002}, in our previous publication we proposed a simpler and possibly more effective strategy for its evaluation, solely based on the (non-linear) DM power spectrum. In this followup we put to use this approach and aim at a critical overview of the different steps and assumptions entering the computation of the EDMF and related uncertainties. We address the dependence of EDMF on cosmological parameters present in a standard cosmology and provide a first assessment of additional errors introduced by simplifications in the cosmology, e.g. neglecting the role of baryons and neutrinos.

In particular, we focus on the critical issue of extrapolation of EDMF below the resolution of N-body simulations. We add the power spectrum measurement from the Millennium Simulation II  \citep[MSII,][]{BoylanKolchinEtal2009} which has a factor of five higher mass resolution than the Millennium Simulation \citep[MS,][]{SpringelEtal2005} used in the previous work, while also adopting a more refined extrapolation scheme. As a consequence, the estimated uncertainty due to the extrapolation shrinks by  one or more orders of magnitude, depending on the redshift, with respect to our previous results. This illustrates the improvement from a better determination of the nonlinear power spectrum from higher resolution simulations. In addition, we derive a large but observational and model-independent upper limit to the redshift integrated EDMF (constraining some of the more aggressive extrapolation prescriptions suggested in the past). This is achieved by taking advantage of the observed scatter in the Type Ia supernovae magnitudes: a quantity  affected by the weak lensing caused by matter perturbations along the line-of-sight and therefore sensitive to the nonlinear matter power spectrum. 

This article is structured as follows. In Section~\ref{sec:formalism} we introduce the central notion of {\it flux multiplier} $\zeta(z)$ and describe its evaluation in terms of the nonlinear matter power spectrum. We discuss some basic theoretical predictions for the power spectrum in the highly nonlinear regime and the extent of state-of-the-art numerical results. Section \ref{sec:systematics} deals with the estimation of the main systematic uncertainties affecting the necessary extrapolation of the power spectrum beyond the limit of simulations resolution. We discuss additional uncertainties related to the difficulty in properly include the effects of baryons and neutrinos and we discuss a possible upper limit from supernovae observations. In Sec.~\ref{sec:cosmouncertainty} we briefly quantify the error budget on the EDMF resulting from cosmological parameters uncertainties in the standard model and from the determination of the extragalactic background light. We provide our conclusions in Section~\ref{sec:conclusions}.

\section{Basic Formalism}
\label{sec:formalism}

\subsection{Gamma-ray flux}
\label{ssec:flux}

For a constant annihilation cross section $\sigv$\footnote{In the case where $\sigv$ is not constant, rather shows a velocity-dependence, $\sigma v$ enters the integrals over the mass distribution and halo profile. The result of averaging over the velocity distribution has been discussed in~\citet{CampbellDuttaKomatsu2010}. We shall not discuss further this complication, since in case the $P-wave$ is dominant, the signal is negligible, the case were Sommerfeld enhancements or resonances are concerned is extremely model dependent and should be discussed case by case.} the extragalactic gamma-ray flux $\phi$ (number of photons per energy interval, unit area, time and solid angle) produced in annihilations of DM particles with mass $m_{DM}$ at a redshift $z$, can be written as \citep[see, e.g.][]{AndoKomatsu2006}
\bea
\phi(E) & = & \frac{c\,\sigv(\Omega_{\rm DM}\,\rho_c)^2}{8\pi\,m_{DM}^2}\,\times\nonumber\\
&  & \int dz \frac{e^{-\tau}(1+z)^3\zeta(z)}{H(z)}\frac{dN(E',z)}{dE'}\Big |_{E' = E (1+z)}\,,\label{finaleq}
\eea 
where $\Omega_{\rm DM}$ is the current DM abundance, $\rho_c$ is the critical energy density, $H$ is the Hubble constant and $dN/dE$ is the spectrum of photons per DM annihilation. The function $\tau$ parametrizes the absorption of photons on the Extragalactic Background Light (EBL) and is further discussed in Section \ref{ssec:statB}. The {\em flux multiplier} is a central quantity of interest in this work and it is defined as
\be
\zeta(z)\equiv\langle \delta^2(z)\rangle\,,
\label{eq:zetaDef}
\ee
namely the variance of dark matter density fluctuations over the sky at a given epoch. 

As noted in  \citet{SerpicoEtal2012} the flux multiplier can be expressed in terms of the {\em non-linear} matter power spectrum $P_{NL}$, i.e. the Fourier transform of the two-point correlation function, equation (\ref{eq:zetaDef}). We have, in fact 
\bea
\zeta(z) & = & \int ^{\kmax} \frac{d\,k}{k}\frac{k^3 P_{NL}(k,z)}{2\pi^2}\nonumber\\
& \equiv & \int ^{\kmax} \frac{d\,k}{k}\Delta_{NL}(k,z)\,,
\label{zetazNL}
\eea 
where we introduced the dimensionless nonlinear power spectrum $\Delta_{NL}(k)= k^3P_{NL}(k)/(2\pi^2)$ (from now on we will drop the explicit redshift dependence of the matter correlators  unless stated otherwise). The above integral is dominated by the high-$k$ behavior (small spatial and mass scales), i.e. the integral depends significantly on the effective upper limit of integration $\kmax$ used, given comparable contributions from each decade in $k$. We will assume, throughout the paper, a sharp cut-off at a given $\kmax$ scale since introducing an arguably more physical function (e.g. an exponential cut-off) would not change our conclusions in any significant way. 

For WIMPs, the cut-off scale depends on collisional damping due to interactions with radiation at early times and to subsequent (collision-less) free-streaming \citep{HofmannSchwarzStocker2001, LoebZaldarriaga2005}, with the latter typically occurring on large scales and therefore constituting the dominant effect \citep{GreenHofmannSchwarz2005}, although acoustic oscillations could also play an important role \citep{LoebZaldarriaga2005}. We note, moreover, that while the collisional damping effect could have determined the cut-off in the initial linear power spectrum, the suppression of perturbations due to free-streaming, i.e. to residual thermal velocities, can be present at much later times and therefore affects even initial stages of nonlinear evolution. Numerical simulations of Earth-mass DM halos, that is close to the cut-off scale, describe the small scale suppression usually in terms of a modified transfer function describing the spectrum of initial fluctuations \citep[see, e.g.][]{DiemandMooreStadel2005}, since a proper description of the phase-space distribution is beyond reach of computational investigations. Some hints on the effects of an initial velocity distribution, however, can be obtained, for instance, from the Warm DM simulations of \citet{VielEtal2012} where a thermal velocity component is added to the particles in the initial conditions. This work shows that the damping scale in the nonlinear WDM matter spectrum presents a small dependence on redshift and the corresponding $k_{\rm cutoff}$ is larger than the one predicted in linear theory. We will not include this dependence on redshift since it cannot be easily extrapolated to the CDM scenario. We will instead assume a constant, comoving $\kmax$, a good approximation in linear theory \citep{GreenHofmannSchwarz2005}.

Previous results on the flux multiplier, based on the HM approach, assumed a cut-off in {\em mass}, but the relation between the DM free-streaming length and the minimal mass of virialized objects that can form is far from obvious. At the heart of the problem lies the assumption that halos closer to the  {\em minimum mass} present structures on much smaller scales, since they are typically described in terms of an NFW \citep{NavarroFrenkWhite1997} profile. However, since the ``size'' of these small halos is determined by physical properties of the dark matter (velocity dispersion, scatterings, \ldots), those halos could show significant departures from the NFW profile and  their very definition of dark matter ``halos'' might become questionable. For example, it was argued in \citet{AnguloWhite2010} based on the extended Press-Schechter formalism that the first virialized objects have mass significantly higher than the one implied by the free-streaming cut-off. Again, some insights can be provided by simulations of WDM or neutrino cosmologies. In the latter case, \cite{Villaescusa-NavarroEtal2013}, using N-body simulations where the neutrino component is described by DM particles but with initial thermal velocities, shows that a small fraction of such particles corresponding to the lowest velocity tail of the distribution, does form virialized objects, but significantly more extended and cored than their purely CDM counterparts (despite forming in the same potential wells). \citet{MaccioEtal2012, MaccioEtal2012ERR}, while adopting a different method to model the thermal velocity distribution of WDM particles, find similar results in terms of more cored halo profiles close to the cut-off mass. In summary, while the link between particle physics properties of DM and the cut-off scale in the linear theory $\kmax$ is well established, it is not clear how it is affected by the non linear evolution. Usual assumption that the smallest halos form at high redshifts (and are therefore highly concentrated) at masses corresponding to the free streaming scale and have universal density profiles might be overreached, in the light of studies mentioned above. Since this plays an important role in the estimate of $\zeta(z)$ we will further discuss this problem in Section~\ref{ssec:resolution}.

\subsection{The nonlinear matter power spectrum}
\label{ssec:matterps}

The evaluation of equation (\ref{eq:zetaDef}) clearly requires knowledge of the matter power spectrum at very small, nonlinear scales, beyond reach of analytical predictions based on perturbation theory (PT). The leading tool to investigate the clustering properties of dark matter deep in the nonlinear regime is given by N-body, numerical simulations. In this regime, in fact, the single-fluid approximation at the basis of perturbative methods is no longer valid and the problem requires the solution of the Vlasov equation in phase-space \citep[see][for a classical review on cosmological PT]{BernardeauEtal2002}. 

However, it is possible to obtain, if not accurate predictions, at least an estimate of the asymptotic behavior of matter correlations from theoretical considerations based on the arguments of self-similar solutions and stable clustering for collision-less particles \citep{DavisPeebles1977}\footnote{See \citet{Peebles1980} and \citet{BernardeauEtal2002} and references therein for a complete introduction.}. Self-similar solutions for the matter correlation functions are possible under the assumptions that there is no characteristic time scale and no characteristic length scale. The first is a good approximation at least during matter domination while the second is valid strictly speaking only for a power-law initial power spectrum, $P_L(k)\sim k^n$; yet,  one can expect it to roughly describe the behavior for large $k$ where $P_L(k)\sim (\ln k)^2/k^3$, corresponding to the suppression of the perturbations on scales that enter the horizon well into radiation domination. The assumption of stable clustering refers instead to the decoupling of collapsed, high-density regions from the Hubble flow. Under these conditions, their physical size is constant in time. This implies that the pairwise velocities of particles within virialized objects cancels, in average, the Hubble expansion. Combined with self-similarity, this assumption leads to exact scale-invariant predictions for the matter correlation function $\xi(x)\sim x^{-\gamma}$ with $\gamma=3(n+3)/(n+5)$ in the case of a power-law initial power spectrum $P_L(k)\sim k^n$ \citep{DavisPeebles1977}. These early results lead the authors of \cite{HamiltonEtal1991} to suggest that the number of pairs within a sphere of radius $x_L$ at early times (i.e. in the linear regime) is constant in time and therefore, $x_L=x\,[1+\bar{\xi}(x)]$, $x$ being the pair separation and $\bar{\xi}(x)$ the averaged correlation function within the sphere at some later time. This establishes an approximate functional relation between late-time correlation function $\xi(x)$ and the early-time correlation function $\xi_L(x_L)$ at some larger scale $x_L$, that is $\xi(x)={\mathcal F}[\xi_L(x_L)]$. This mapping has been extended in \citet{PeacockDodds1994} to the power spectrum, providing an ansatz for the nonlinear power spectrum $\Delta(k)$ which can be expressed as a function of the linear one $\Delta_L(k)$ as 
\be\label{eq:PD94}
\Delta(k)={\mathcal F}[\Delta_L(k_L)]\,,
\ee
with $k=[1+\Delta(k)]\,k_L$. In the limit of small-scales, in particular, stable clustering implies ${\mathcal F}(\Delta)\simeq \Delta^{3/2}$. Such simple relation, which assumes as well $k\sim k_L$, is all that we will make use of in the following, since we are essentially interested in the asymptotic behaviour of the power spectrum and since at very large $k$ the linear spectral index tends to a constant. We notice, however, that stable clustering considerations can be useful also in the HM approach, as recently shown by \citet{ZavalaAfshordi2013B}.

Equation~(\ref{eq:PD94}), in practice, is a quite crude approximation \citep{SmithEtal2003} and, in particular, is not expected to be a universal relation. Still, it constituted the basis for several fitting functions with the map ${\mathcal F}$ now depending on cosmological parameters (the spectral index at the transition between linear and nonlinear scales, in the first place) and on a few free parameters to be determined by comparison with simulations. The most popular is perhaps the \texttt{halofit} formula of \citet{SmithEtal2003} (HF, from now on) recently updated with a revised version (denoted here by RHF) by \cite{TakahashiEtal2012}\footnote{An alternative fitting of the HF formula, but less extensively tested, has been proposed by \citet{InoueTakahashi2012} to describe simulations results up to $k\simeq 320\kMpc$.}. We notice that the emphasis in these works---particularly in the latter one---has been placed on the accuracy of the fit, with respect to a relatively large region in parameter space for $\Lambda$CDM models and some extensions, over the range of scales currently probed by simulations. This implies that the asymptotic behavior predicted by stable clustering for values of $k$ beyond the simulation resolution is not, as we will see, rigorously reproduced. 

As already mentioned, the primary tool to tackle nonlinear clustering is N-body simulations. For the problem at hand we need large-volume simulations, i.e. simulations run over a volume that represents a ``fair sample'' of the Universe. In practical terms, one can require, for instance, that the fundamental frequency of the simulation box, $k_f\equiv 2\pi/L$, $L$ being the linear size of the box, is still in linear or mildly nonlinear regime at $z=0$. Such requirement is approximately equivalent to ensure that the size of simulated volume is much larger than the typical size of collapsed structures, i.e. dark matter halos. A good, state-of-the-art, example of cosmological simulations that allow to explore the nonlinear regime of the matter power spectrum is given by the Millennium and Millennium II simulations \citep{SpringelEtal2005, BoylanKolchinEtal2009}, characterized by a mass resolution of $8.6\times 10^8$ and $6.9\times 10^6\Ms$, respectively. The comparison between these two runs, as we will see, illustrates as well the improvement on the power spectrum determination achievable by a significant increase in resolution. 

We should notice that the Millennium simulations are not the highest-resolution N-Body simulations available. For instance, the Via Lactea \citep{DiemandKuhlenMadau2007} and Aquarius \citep{SpringelEtal2008} projects explore the evolution of Milky Way-size dark matter halos with much greater resolution. However, the limited simulated volume represents a specific ``environment'' of the Universe and a direct measurement of the matter power spectrum is not directly useful to our purposes.

\begin{figure*}
{\includegraphics[width=0.9\textwidth]{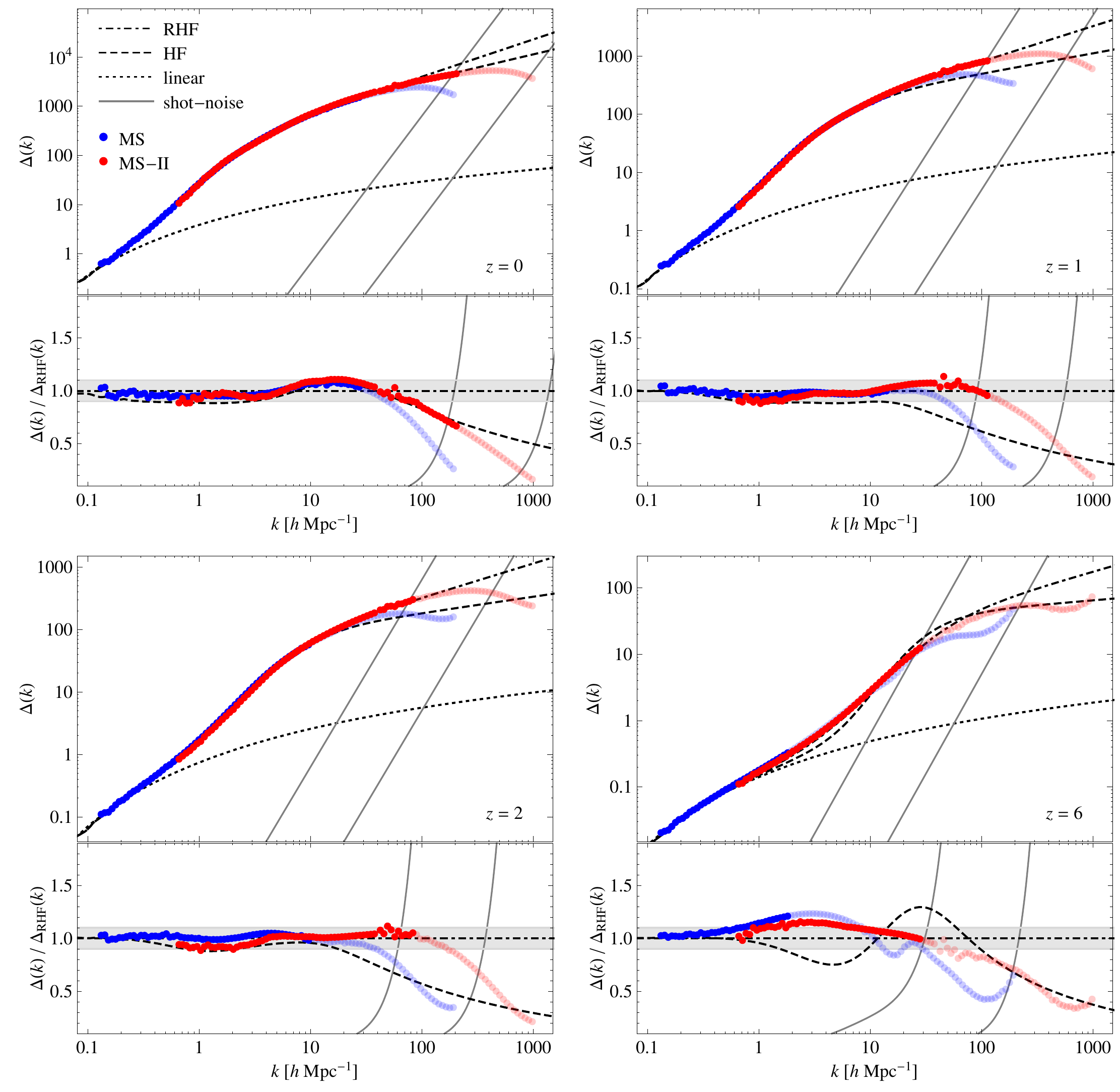}} 
\caption{Measurements of the nonlinear matter power spectrum in the MS ({\em blue data points})
 and MSII ({\em red data points}), respectively~\citep{SpringelEtal2005, BoylanKolchinEtal2009}. Points with lighter shade correspond to $k>k_{1\%}$, i.e. scales where the shot-noise contribution exceeds 1\%. The numerical results are compared to the predictions of the {\halofit} code (HF, {\em dashed, black curve}) and of its revised version of \citet{TakahashiEtal2012} (RHF, {\em dot-dashed, black curve}). The dotted black curve represents the linear matter power spectrum while while the gray lines provide the shot-noise contribution (removed from the measured quantities). Different panels correspond to redshift $z=0$, 1, 2 and 6, as shown, with the lower half of each panel showing the ratio to the RHF prediction. Since the output snapshots of MS and MSII do not exactly share the same redshift values, the MS data points have been rescaled by the square of the linear growth factor for clarity. This does not ensure a perfect match in the nonlinear regime, but differences are negligible for our purposes.}
\label{fig:MSps}
\end{figure*}

In Fig.~\ref{fig:MSps} we show measurements of the nonlinear matter power spectrum in the MS ({\em blue data points}) and MSII ({\em red data points}). The numerical results are compared to the predictions of the {\halofit} code\footnote{All evaluations of the original \texttt{halofit} code include the small-scale correction suggested on John Peacock's webpage \href{http://www.roe.ac.uk/~jap/haloes/}{http://www.roe.ac.uk/~jap/haloes/}. This is not implemented for the RHF predictions.} (HF, {\em dashed, black curve}) and of its revised version of \cite{TakahashiEtal2012} (RHF, {\em dot-dashed, black curve}). The dotted black curves represent the linear matter power spectrum while while the gray lines provide the shot-noise contribution, removed from the measured quantities. Different panels correspond to redshift $z=0$, 1, 2 and 6, as shown, while the lower half of each panel shows the ratio to the RHF prediction for the different quantities.

The determination of the power spectrum at the smallest scales is significantly affected by the resolution of the simulation, even on scales where the shot-noise contribution is subdominant. This can be clearly seen by the comparison of the MS and MSII results. We notice that a good agreement between the two runs is present up to scales where shot-noise contributes a few percent or less. At redshift $z=0$, a 1\% shot-noise level corresponds to a value of $k=29$ and $210\kMpc$ for MS and MSII, respectively. For a given simulation and for a given redshift we will denote such scale as $k_{1\%}$. The scale is reduced at larger redshift, due to the lower value of the matter power spectrum itself. 

The RHF fitting function has been determined from the analysis of a set of simulations on scales $k\le30\Mpc$ and up to redshift 10. They report a precision of 5\% at large scales ($k<1\Mpc$) and 10\% at smaller scales and low redshift ($0<k<10\Mpc$ for $0<z<3$), confirmed by our own comparison with the Millennium simulations measurements. As we will discuss in detail in the next section, the accuracy of the RHF or HF fitting formulas over the range of scales probed by simulation is more than sufficient for our purposes, while the real problem is the extrapolation of such expressions to larger $k$. 

Before ending this section, we notice that stable clustering is not the only analytical tool to investigate the highly nonlinear regime of the matter power spectrum. An expression for the nonlinear power spectrum can, in fact, be obtained as well in the framework of the Halo Model itself \citep[see][for a review]{CooraySheth2002} and in terms of the same ingredients required for the determination of the flux multiplier $\zeta(z)$. For example, a recent application of such a computation to cross-correlations between gamma-ray anisotropies and cosmic shear has been done by \citet{CameraEtal2013}. A comparison of such prediction with measurement from N-body  simulations could constitute a useful check, particularly since power spectrum measurements account for the full nonlinear density distribution, beyond the simplifications and approximations of the HM approach.

\section{Systematic uncertainties on the flux multiplier $\zeta$}
\label{sec:systematics}

From the cosmological point of view, the whole problem of computing $\zeta(z)$ reduces to determining the power spectrum at very small scales. The evaluation of the dark matter annihilation flux will therefore suffer, in the first place, from the significant systematic uncertainty associated with our poor knowledge of the properties of matter perturbations at very small scales. In addition to the simple lack of numerical results due to the simulations limited resolution, other sources of uncertainty are given by the effects of baryons and neutrinos. The former in particular constitute an outstanding problem. Hydrodynamical simulations accounting for baryonic physics are computationally more demanding. To this one should add that the numerical models themselves are relative simple, compared to the complexity of the problem. 

In this section we will attempt to provide, in the first place, an estimate of the uncertainty related to the extrapolation of current numerical results for the (collision-less) dark matter power spectrum to extremely small scales. We will also discuss how introducing realistic descriptions of baryonic and neutrino effect might alter the picture. We consider all such uncertainties as ``systematic'' as they depend on our ignorance on structure growth at small scales, possibly resulting in incorrect extrapolations. Of course, one can consider as well ``statistical'' uncertainties, due to our limited knowledge of cosmological parameters, even assuming the correctness of the underlying model. These ``minor'' uncertainties will be briefly discussed in section \ref{sec:cosmouncertainty}, together with remaining uncertainties related to the modeling of the photon absorption on the EBL light.

\subsection{Limited resolution} 
\label{ssec:resolution}

In order to estimate the extrapolation error, we first make a very simple assumption inspired by direct inspection of the simulations. We notice, in the first place, that the nonlinear adimensional power spectrum $\Delta(k,z)$ is a non-decreasing function of $k$. In addition, its second derivative is negative at sufficiently large scales, i.e. beyond the transition between linear and nonlinear regimes. This is evident in the range probed by the MS and MSII measurements shown in Fig.~\ref{fig:MSps} but it also a generic prediction of stable clustering and self-similar solutions. We will assume these functional properties to hold at all smaller scales. Finally, concerning the evolution in redshift, we shall impose that the power spectrum at each comoving scale $k$ is a non-decreasing function of time. This is again consistent both with what is inferred from linear theory and via simulations, in their resolved scales. We shall demand that this is true also at non-linear scales currently unresolved.

Let us now define $k_\star$ as the largest wavenumber at which numerical results can be trusted. Then, assuming the extrapolation to be scale-invariant, we can reasonably expect the true value of $\Delta_{NL}(k,z)$, for $k>k_\star$, to be bounded as $\Delta_{\rm min}(k,z)\le \Delta(k,z)\le \Delta_{\rm max}(k,z)$ with
\be
\label{eq:extrap1}
\Delta_{\rm min}(k,z)  =   \Delta(k_\star,z)\,,
\ee
and
\bdm
\Delta_{\rm max}(k,z)  = 
\edm
\be
\label{eq:extrap2} 
 {\rm Min}\left[\Delta(k_\star,z)\left(\frac{k}{k_\star}\right)^{n_{s,{\rm eff}}(k_\star)+3}\!, \Delta_{\rm max}(k,z')\right]\forall\, z'\!<\!z\,,
\ee
where 
\be\label{eq:ns}
n_{s,{\rm eff}}(k)\equiv \frac{\de \ln P(k)}{\de\ln k}\,,
\ee
is the effective spectral index of the nonlinear power spectrum at scale $k$. The upper limit given by the first expression in square brackets therefore preserves the continuity of $\Delta(k)$ and its derivative at $k_\star$. The minimum condition enforces the ``non-negative time evolution'' condition $\partial\Delta(k,z)/\partial z\le 0$.

\begin{figure*} 
{\includegraphics[width=0.9\textwidth]{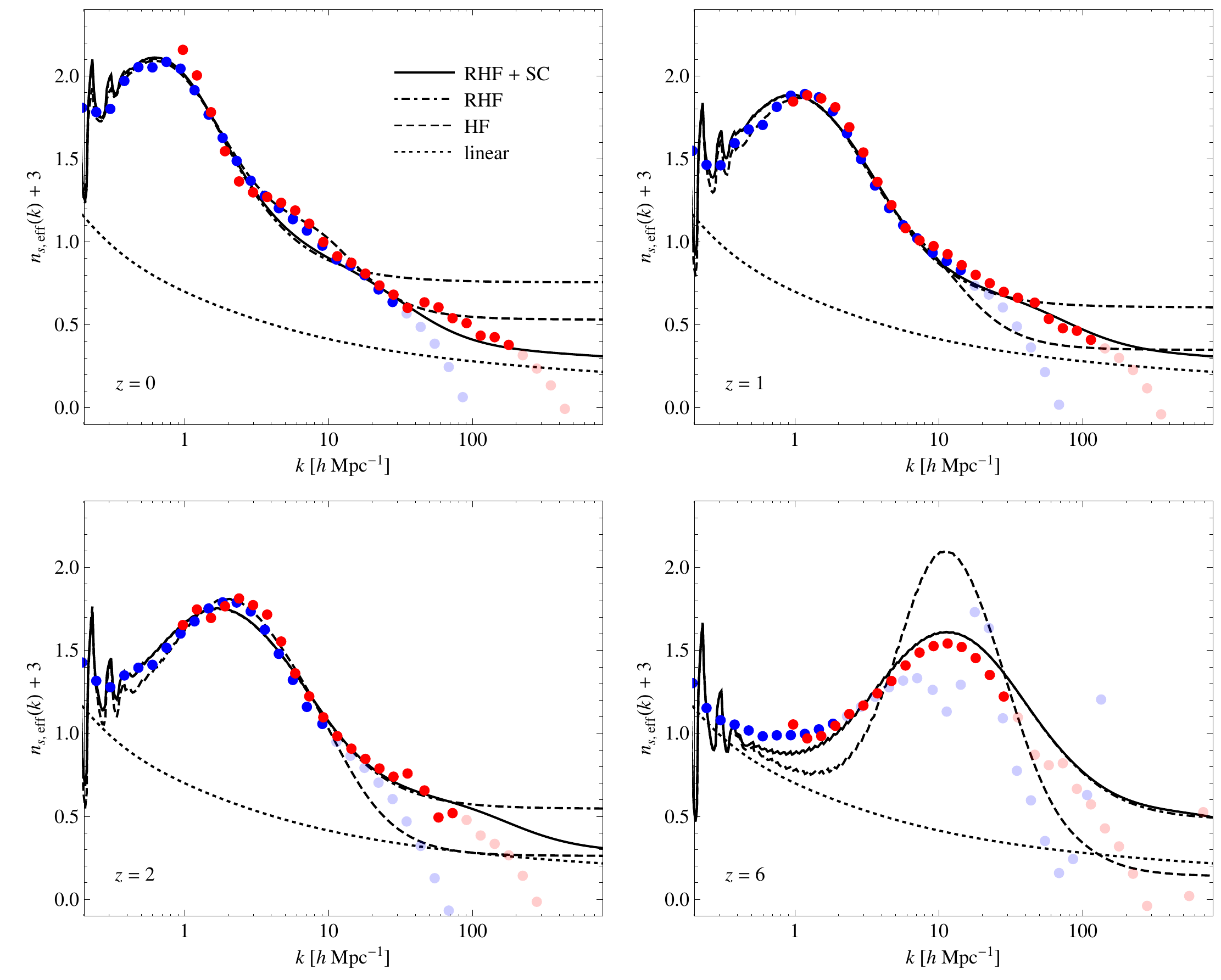}} 
\caption{Effective spectral index $n_{s,{\rm eff}}(k)$, equation (\ref{eq:ns}), derived from the measurements of the nonlinear matter power spectrum in the Millennium ({\em blue data points}) and Millennium II ({\em red data points}) simulations \citep{SpringelEtal2005, BoylanKolchinEtal2009}. The numerical results are compared to the predictions of the HF formula ({\em dashed, black curve}), to its revised version of \citet{TakahashiEtal2012} ({\em dot-dashed, black curve}) and to the expression of equation~(\ref{eq:RHFSC}) (RHF+SC, {\em black, continuous curve}). The dotted line shows the effective spectral index for the linear power spectrum \citep[without baryonic features, from the fit of][]{EisensteinHu1998}. Vertical gray lines indicate the scales at which the shot-noise contribution reaches the 1\% level to the measured power spectra. Beyond the scale, the data points are marked with lighter colors since they are discarded in our estimates. Different panels correspond to redshift $z=0$, 1, 2 and 6, as shown.
}
\label{fig:ns}
\end{figure*}

In order to estimate these limiting values we need to define the scale $k_\star$. A possible definition can be given in terms of the level of shot-noise contribution to the power spectrum, since this is a direct manifestation of the finite resolution. We can set, for instance, $k_\star$ equal to the scale $k_{1\%}$ where the shot-noise contributes by 1\% to the measured, total power spectrum before shot-noise correction. Clearly this choice is rather arbitrary. However, we can further motivate the specific value of 1\% for the level of the shot-noise contribution, by directly comparing measurements from MS and MSII. In this respect, the relevant quantity is the effective spectral index, equation (\ref{eq:ns}), measured in the two simulations since this quantity directly affects the estimate of the upper bound of the extrapolation more than the power spectrum amplitude itself. Such comparison is shown in Fig.~\ref{fig:ns}. Again MS and MSII data points are shown by blue and red data points, respectively, while data points in lighter shade correspond to wave numbers beyond $k_{1\%}$. For $k<k_\star=k_{1\%}$, where the results from the two simulations overlap, they provide consistent results. 

With some additional but rather simple assumptions, it is possible to provide tighter bounds for the nonlinear power spectrum, with respect to those of equations (\ref{eq:extrap1}) and (\ref{eq:extrap2}). Regarding the lower bound, we can expect from stable clustering as well as from what shown in figure \ref{fig:ns}, that the nonlinear effective spectral index is larger or equal to the spectral index predicted by linear theory. This can lead to the more restrictive definition of $\Delta_{\rm min}$ given by
\be\label{eq:extrap1C}
\Delta_{\rm min}(k,z)  = \Delta(k_\star,z)\left(\frac{k}{k_\star}\right)^{n_{s,L}(k_\star)+3}\,,
\ee
where $n_{s,L}$ is 
\be\label{eq:nsL}
n_{s,L}(k)\equiv \frac{\de \ln P_L(k)}{\de\ln k}\,,
\ee
$P_L(k)$ being the linear power spectrum. For the upper bound, we can assume that nonlinear growth at fixed $k$ will be larger than the one predicted by the linear growth factor $D(z)$, leading to  
\bdm
\Delta_{\rm max}(k,z) =
\edm
\bdm
 {\rm Min}\left[\Delta(k_\star,z)\left(\frac{k}{k_\star}\right)^{n_{s,{\rm eff}}(k_\star)+3},\,\frac{D^2(z)}{D^2(z')}\Delta_{\rm max}(k,z')\right]
\edm
\be\label{eq:extrap2C} 
\forall\, z'<z\,,
\ee
where we rescale the upper bound at lower redshift $z'$ to the redshift of interest $z$ by the linear factor $D^2(z)/D^2(z')$.

Finally, along with the uncertainty estimates described above, we would like to provide as well a ``best guess'' for the nonlinear power spectrum extrapolation. We assume in the first place that the RHF formula is accurate enough over the range of scale probed by simulations. For large $k$ we impose, for simplicity, the most crude approximation of stable clustering predictions, that is $\Delta(k)\sim \Delta_L^{3/2}(k)$. In order to ensure a smooth transition we consider the expression
\bea\label{eq:RHFSC}
\Delta(k) & = &  \frac1{1+(k/k_t)^\alpha}\Delta_{\rm RHF}(k)\nonumber\\
& & +\frac{(k/k_t)^\alpha}{1+(k/k_t)^\alpha}\Delta_{\rm RHF}(k_t)\left[\frac{\Delta_L(k)}{\Delta_L(k_t)}\right]^{3/2}\,,
\eea
where $\alpha=1.5$ and $k_t(z)=100 \knl(z)$, $\knl(z)$ being the ``nonlinear'' scale determined by the HF (and RHF) algorithm from the equality $\sigma^2(1/\knl)=1$ with
\be
\sigma^2(R)=\int \de\ln k\,\Delta_L(k)\,e^{-k^2\,R^2}\,.
\ee
We will refer to this prescription as the Revised \texttt{halofit} plus Stable Clustering (RHF-SC). Clearly, both the parameter $\alpha$ and the transition scale $k_t$ could be obtained directly from fits to the numerical results along with the other RHF parameters. Our {\em ad-hoc} choice, however, is sufficient for our purposes. We just notice that the values of $k_t(z)$ so defined are typically larger than the scales $k_{1\%}$ discussed in the previous section, particularly at large redshift. In those cases, part of the extrapolation, for $k_{1\%}<k<k_t$, is given by the RHF formula itself.   

Fig.~\ref{fig:ns} show as well the predictions for the effective spectral index of the HF ({\em dashed curves}) and RHF ({\em dot-dashed curves}) formul\ae, along with those of the corrected expression of equation (\ref{eq:RHFSC}) (RHF+SC, {\em continuous curve}) and for the linear power spectrum\footnote{We adopt here for its ease of use the fitting formula for the linear transfer function of \citet{EisensteinHu1998}.} ({\em dotted curve}). Interestingly, the correction to RHF motivated by stable clustering, improves the prediction for the MSII measurements at $z=0$ in the range $30<k<200\kMpc$, beyond the limit of validity of the RHF formula. 

\begin{figure*}
{\includegraphics[width=0.9\textwidth]{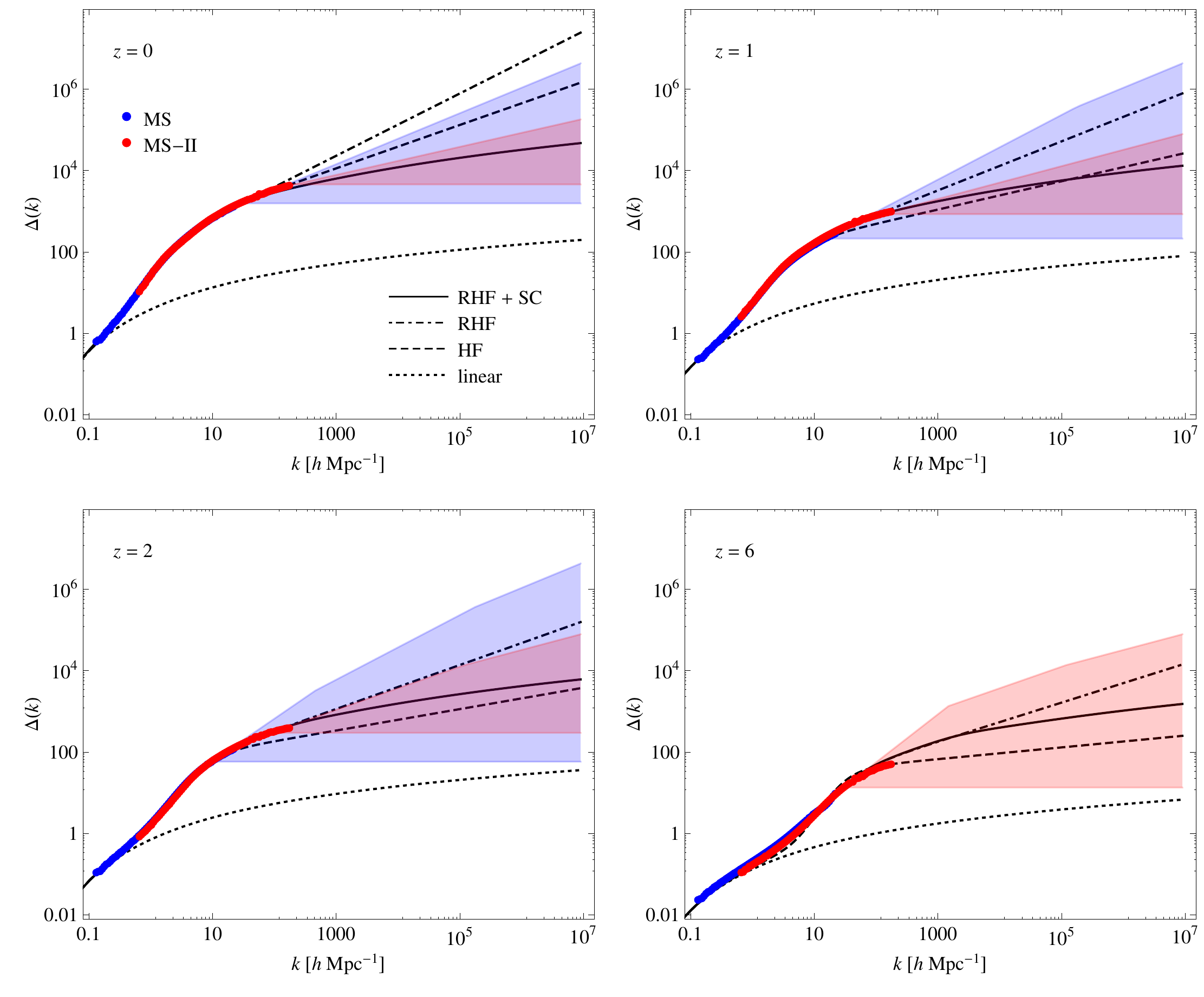}} 
\caption{Extrapolations of $\Delta(k)$ to small scales. All curves are the same as in Fig.~\ref{fig:MSps}, with data points shown only up to $k_\star=k_{1\%}$ (see text for explanation). In addition, the blue shaded area shows the allowed range for the nonlinear power spectrum defined by equations (\ref{eq:extrap1}) and (\ref{eq:extrap2}) and estimated from the MS data alone up to $k_\star$. The red area is corresponds instead to the allowed range estimated from MSII data. The black, continuous curve shows the prediction of RHF corrected at large scales according to equation (\ref{eq:RHFSC}).  Different panels correspond to redshift $z=0$, 1, 2 and 6, as shown. For $z=6$ only the MSII extrapolation is considered, since the MS data do not cover sufficiently well nonlinear scales.}
\label{fig:extrap}
\end{figure*}

\begin{figure*}
{\includegraphics[width=0.96\textwidth]{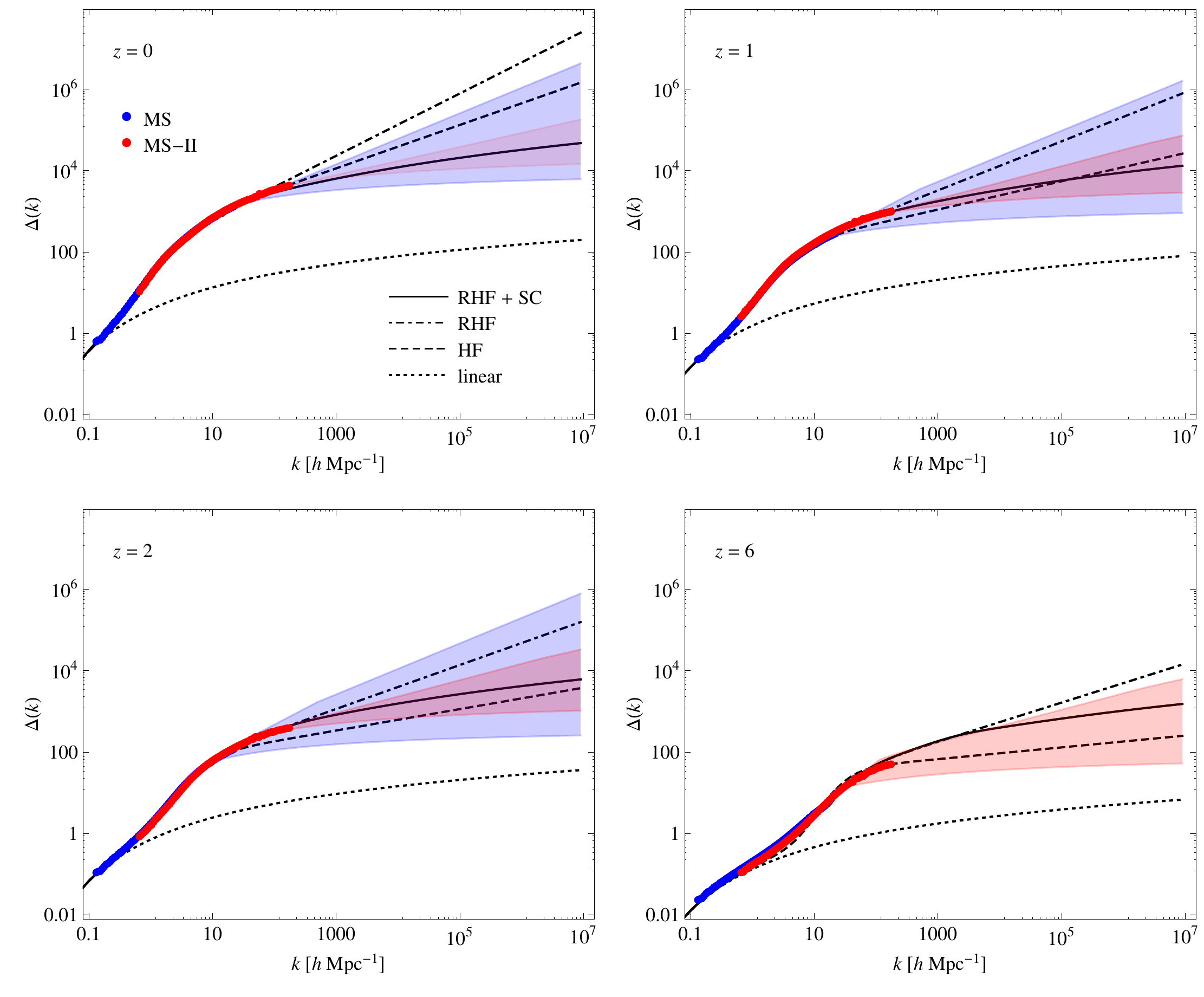}} 
\caption{Same as Fig~\ref{fig:extrap} but for the limits defined in equations (\ref{eq:extrap1C}) and (\ref{eq:extrap2C}).}
\label{fig:extrapC}
\end{figure*}

The extrapolations of $\Delta(k)$ defined in equations (\ref{eq:extrap1}) and (\ref{eq:extrap2}) are shown in Fig.~\ref{fig:extrap}, together with the data points of MS and MSII up to $k_\star=k_{1\%}$. The shaded areas enclose the allowed regions derived from the MS data ({\em blue}) and MSII data ({\em red}). At redshift $z=6$ only the MSII extrapolations are considered because the MS data do not  cover sufficiently well the nonlinear regime: as can be seen from Fig.~\ref{fig:ns} the corresponding $k_\star$ falls before the range where $\neff(k)$ is decreasing and we cannot therefore expect the upper limit extrapolation to be a reasonable estimate. Fig.~\ref{fig:extrapC} shows instead the limits defined in equations (\ref{eq:extrap1C}) and (\ref{eq:extrap2C}). The upper limit at $z=0$ is the same, by definition, for both cases.

\begin{table*}
\renewcommand{\arraystretch}{1.2} \renewcommand{\tabcolsep}{0.1cm}
\begin{tabular}{||c||c|cc|c||c|cc|c||}
\hline
 & \multicolumn{8}{c||}{Enhancement factor $\zeta\times 10^{-3}$} \\
\cline{2-9}
 & \multicolumn{4}{|c||}{$k_{\rm max}=100\kMpc$}    & \multicolumn{4}{c||}{$k_{\rm max}= 10^6\kMpc$} \\ 
 \cline{2-9}
$z$ & MS I & MS II & & RHF+SC &   MS I & MS II & & RHF+SC  \\
\hline\hline
 \multicolumn{9}{||l||}{{\em Conservative bounds}, equations (\ref{eq:extrap1}) and (\ref{eq:extrap2})}\\
\hline
0.0 & 3.6~/~4.6 & 4.7& (0.1) & 4.4   & 18~/~1700 & 46~/~240& (1.3) & 130 \\
\hline
1.0 & 0.65~/~1.2 & 1.1& (0.24) & 1.1  & 2.8~/~1700 & 9~/~79& (1.8) & 36 \\
\hline
2.0 & 0.18~/~0.62 & 0.44~/~0.45& (0.54) & 0.45   & 0.71~/~1600 & 3.2~/~72 & (2.0) & 17 \\
\hline
6.0 & 0.0016~/~0.021 & 0.026~/~0.046& (0.88) & 0.045 & - & 0.15~/~71 & (-) & 4.1\\
\hline
\hline
 \multicolumn{9}{||l||}{{\em Tighter bounds}, equations (\ref{eq:extrap1C}) and (\ref{eq:extrap2C})}\\
\hline
0.0 & 4.0~/~4.6 & 4.7& (0.058) & 4.4   & 42~/~1700 & 84~/~240& (1.2) & 130 \\
\hline
1.0 & 0.8~/~1.2 & 1.1& (0.18) & 1.1  & 7.0~/~690 & 17~/~79& (1.3) & 36 \\
\hline
2.0 & 0.24~/~0.53 & 0.44~/~0.45& (0.34) & 0.45   & 1.9~/~310 & 6.5~/~36 & (1.5) & 17 \\
\hline
6.0 & 0.0031~/~0.021 & 0.029~/~0.046& (0.64) & 0.045 & - & 0.35~/~6.8 & (-) & 4.1\\
\hline
\end{tabular}
\caption{Allowed range for the enhancement factor $\zeta$ (divided by 1,000) at four different redshifts. Values are estimated from the extrapolations in equations (\ref{eq:extrap1}) and (\ref{eq:extrap2}), as well as those of equations (\ref{eq:extrap1C}) and (\ref{eq:extrap2C}), of the MS and MSII data together with the prediction from the revised \texttt{halofit} formula corrected at large scales according to equation (\ref{eq:RHFSC}) (RHF+SC). The value of $k_{\star}= k_{1\%}$ is assumed while two values for the cut-off $k_{\rm max}$ in the integral of equation (\ref{zetazNL}) are considered. The lower one, $k_{\rm max}=100\kMpc$ corresponds to scales currently probed by numerical simulations and therefore provides a lower bound to enhancement factor. The higher value, $k_{\rm max}=10^6\kMpc$, falls instead in the typical range of values expected for the dark matter free-streaming cut-off. The number in parenthesis provides the improvement in {\it orders of magnitude} from the MS to the MSII extrapolations, that is $\log_{10}(\zeta_{\rm max}/\zeta_{\rm min})_{\rm MS}-\log_{10}(\zeta_{\rm max}/\zeta_{\rm min})_{\rm MSII}$. For the MSII case and $k_{\rm max}=100\kMpc$ there is no extrapolation and therefore no uncertainty since $k_{\rm max}<k_{\star}$. }\label{tab:zeta}
\end{table*}

We notice, in the first place, that the increase in resolution provided by MSII with respect to MS corresponds to a significant reduction in the uncertainty estimated by our simple extrapolation. This is quantified in Table~\ref{tab:zeta}, where the allowed range for the enhancement factor, $\zeta_{\rm max}$ / $\zeta_{\rm min}$ (divided by 1,000) at four different redshifts is estimated from the extrapolations of the MS and MSII data. The improved resolution is responsible for a reduction in the estimated allowed region by  about two orders of magnitude at $z=2$ for $k_{\rm max}=10^6\kMpc$ for the conservative limits, slightly less for the tighter ones. Both our definitions of the upper and lower limits for $\Delta(k)$ imply that the relative uncertainty at $k>k_\star$ grows with redshift, since the value of $k_\star=k_{1\%}$ is lower at higher redshift and, in turn, the value of $\neff(k_\star)$ larger. Since at larger redshift we are probing a narrower range of scales, this choice enforces more conservative results at early time. This is desirable: the effect of a finite mass resolution is larger at large redshift where structure are less evolved and where it is therefore more challenging to quantify their properties. In fact, this is true as well for the typical ingredients required by the Halo Model approach, such as the mass function, the halo profile, etc. However, the dependence on redshift of the uncertainty in the determination of such quantities is not accounted for (instantaneous virialization and convergence to asymptotic universal halo profile is for example assumed).  Note however how the minimum condition enforced via equation (\ref{eq:extrap2}) prevents the error to grow too much, with a moderating effect that is more pronounced at high $z$ and high $k$.

Fig.s~\ref{fig:extrap} and ~\ref{fig:extrapC} show as well, for comparison, the extrapolation of the HF and RHF fitting formulas, together with the corrected version of equation (\ref{eq:zetaDef}) enforcing the stable clustering prediction. Both the extrapolated values of HF and RHF exceed the bounds derived from the simulations. This is not surprising since, as mentioned before, the large-$k$ asymptotic behavior has not been considered in the fitting procedure. On the other hand, the  stable clustering assumption provides a ``best guess'' extrapolation that nicely falls within the estimated limits, both from MS and MSII, for all redshifts considered, even in the case of the tighter aggressive limits of equations (\ref{eq:extrap1C}) and (\ref{eq:extrap2C}). This is evident as well confronting the values obtained for $\zeta(z)$ with the allowed interval as reported in Table~\ref{tab:zeta}. 

\begin{figure*} 
{\includegraphics[width=0.9\textwidth]{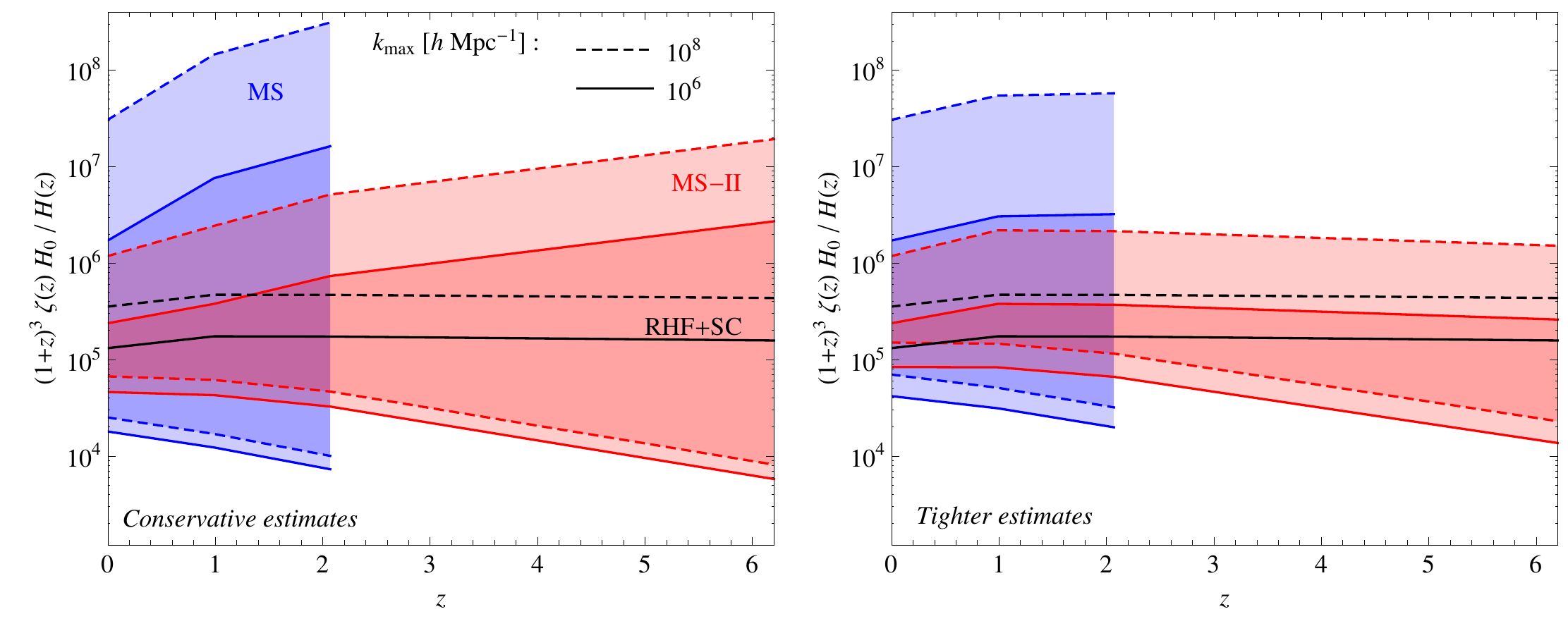}}
\caption{Estimated uncertainty on the quantity $(1+z)^3\,\zeta(z)\,H_0\,/\,H(z)$, equation (\ref{finaleq}), as a function of redshift from the extrapolated MS data ({\em blue regions}) and MSII data ({\em red regions}) and the best guess prediction of equation (\ref{eq:RHFSC}), that is from the RHF fitting formula plus stable clustering extrapolation (RHF+SC, {\em black curves}). Results are shown for the two values $\kmax=10^6$ and $10^8\kMpc$ ({\em continuous and dashed curves, respectively}). The left panel assumes the more conservative bounds of equations (\ref{eq:extrap1}) and (\ref{eq:extrap2}) while the right panel assumes equations (\ref{eq:extrap1C}) and (\ref{eq:extrap2C}). }
\label{fig:zetaMS}
\end{figure*}

These results are visualized as well in Fig.~\ref{fig:zetaMS} where the uncertainty on the dimensionless combination $(1+z)^3\,\zeta(z)\,H_0/\,H(z)$ estimated from the extrapolated MS data ({\em blue regions}) and MSII data ({\em red regions}) is shown as a function of redshift. Black curves correspond to the RHF+SC prediction. Two different values for the integration cut-off are considered, $k_{\rm max}=10^6$ and $10^8\kMpc$ ({\em continuous and dashed curves, respectively}). All extrapolations assume $k_\star=k_{1\%}$. The left panel assumes the more conservative bounds of equations (\ref{eq:extrap1}) and (\ref{eq:extrap2}) while the right panel assumes equations (\ref{eq:extrap1C}) and (\ref{eq:extrap2C}). Clearly the lower bounds are not affected much by the two orders of magnitude difference in the cut-off assumed here, while the upper bounds change by up to about a factor of ten, depending on the redshift, in the conservative extrapolations case. Notice that we limit the plots to the four outputs available, $z=0$, 1, 2 and 6 and that we have no upper bounds estimated from MS at redshfit $z=6$, so we stop at $z=2$. The estimated uncertainties obviously depend as well on the choice of $k_\star$, the starting point for the extrapolations: choosing $k_\star=k_{0.1\%}$ would have lead to larger bounds by roughly an order of magnitude at all redshifts. Such a choice, however, would be probably too conservative as it would ignore a large range of scales well probed by simulations, as one can estimate from the comparison between MS and MSII. 

\begin{figure*}
{\includegraphics[width=0.9\textwidth]{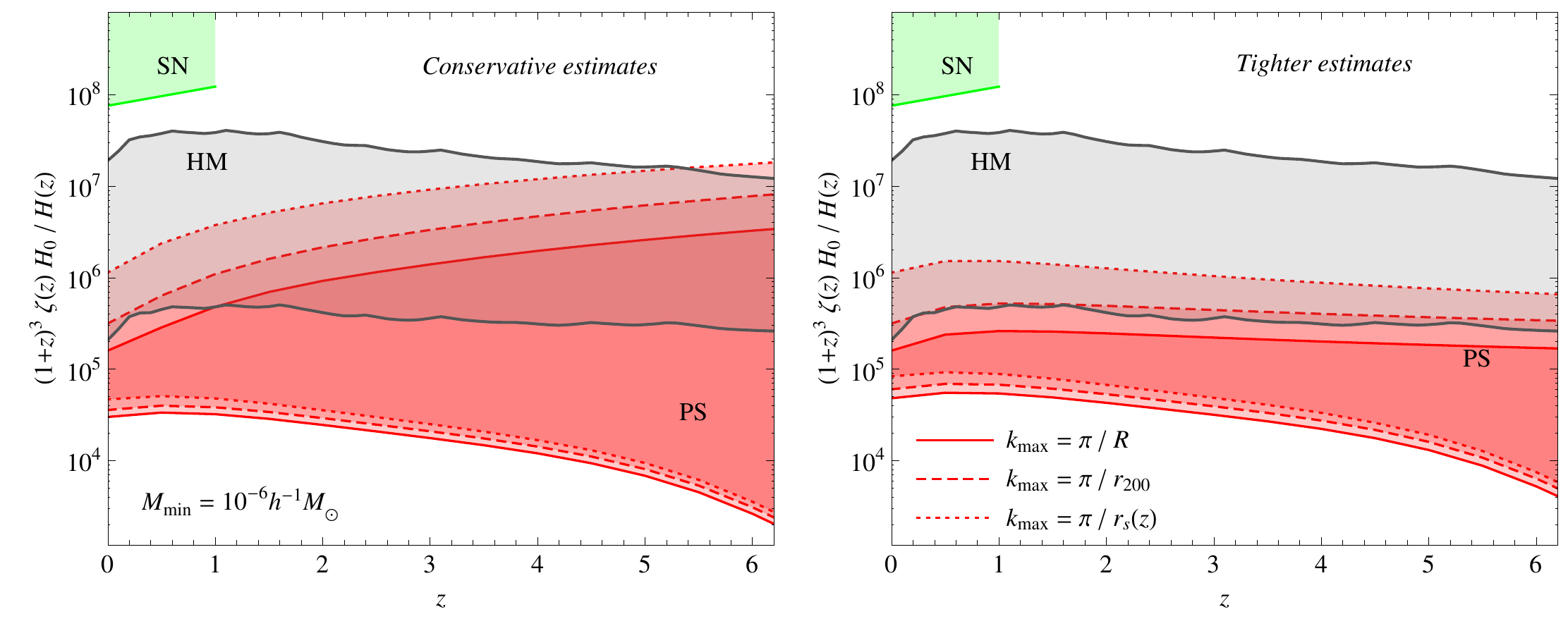}} 
\caption{Comparison of the quantity $(1+z)^3\,\zeta(z)\,H_0\,/\,H(z)$, as a function of redshift, evaluated in the HM approach by \citet{ZavalaSpringelBoylan-Kolchin2010,AbdoEtal2010B} for $M_{\rm min}=10^{-6}\Ms$ ({\em gray region bounded by continuous curves}) to the results from the power spectrum approach proposed in this work. Three different, possible values of $\kmax$, both corresponding to  $M_{\rm min}=10^{-6}\Ms$ are considered. The first defines $\kmax=\pi/R$, with $R=[3M_{\rm min}/(4\pi\bar{\rho}_m)]^{1/3}$ ({\em continuous curves}) and therefore corresponding to the physical size of a perturbation of mass $M_{\rm min}$ in the initial density field; in the second case we assume  $\kmax=\pi/r_{\rm 200}$ ({\em dashed curves}) , $r_{\rm 200}$ being the size of a collapse spherical overdensity of mass $M_{\rm 200}=M_{\rm min}$ where the mean density is 200 times the critical density; in the third case, we use  $\kmax=\pi/r_s$, with $r_s$ the (now redshift dependent) scale radius of the NFW profile (see text for explanation). The green area denotes the region probed by the effects of weak lensing by nonlinear perturbations on the variance of SN magnitudes (see Sec.~\ref{ssec:SN}). All extrapolations assume $k_\star=k_{1\%}$. The left panel assumes the more conservative bounds of equations (\ref{eq:extrap1}) and (\ref{eq:extrap2}) while the right panel assumes equations (\ref{eq:extrap1C}) and (\ref{eq:extrap2C}).}
\label{fig:zeta}
\end{figure*}

Fig.~\ref{fig:zeta} attempts a comparison of our results based on the power spectrum extrapolation with the results obtained in the Halo Model approach by \citet{ZavalaSpringelBoylan-Kolchin2010} (and presented in \cite{AbdoEtal2010B}) using the MSII simulation to predict the EDMF flux ({\em gray region, bounded by dotted curves}). For a further example of comparison between HM and PS predictions see \citet{NgEtal2013}. The HM estimation of the allowed region assumes power-law extrapolations of the gamma-ray luminosity of halos up to the minimal halo mass cutoff $M_{\rm min}=10^{-6}\Ms$. In addition,  the luminosity contribution from subhalos was also extrapolated both in normalization and slope with a power-law, down to the same scale, for each host halo mass. The most conservative choice of extrapolation parameters then defines the lower limit while the most aggressive extrapolation (of all quantities involved) defines the upper limit. 

In order to provide a more realistic prediction for the flux multiplier and its uncertainty, our power spectrum calculations assume, for Fig.~\ref{fig:zeta}, a $\Lambda$CDM comsology defined by the Planck best-fit parameters \citep{Planck2013parameters}. In this case we simply assume the RHF formula to accurately describe the nonlinear power spectrum up to $k_\star=k_{1\%}$ where $k_{1\%}$ is defined assuming only the shot-noise level of MS-II. The extrapolations are defined is the same way as before. Of course, this allows us to provide estimates for the uncertainty in $\zeta$ at any redshift, without beeing limited to the available MS outputs. 

To enable a comparison between the HM and PS methods, the cut-off $\kmax$ in the power spectrum evaluation should be chosen to reproduce the results obtain in the HM approach assuming $M_{\rm min}=10^{-6}\Ms$.  As mentioned above, while the definition of $\kmax$ as a function of the DM free-streaming length in the \emph{linear} regime is unambiguous, the definition of a minimal halo (and subhalo) mass as well as density profiles within the smallest halos is not. The choice of $M_{\rm min}=10^{-6}\Ms$ assumed in \citet{ZavalaSpringelBoylan-Kolchin2010} as a typical cut-off mass, is motivated by the results of \citet{Bringmann2009}, where a minimal protohalo mass is associated to a free-streaming wavelength $k_{\rm fs}$ simply as
\be
M_{\rm fs}=\frac{4\pi}3\,\bar{\rho}\,\left(\frac{\pi}{k_{\rm fs}}\right)^3\,.
\ee
Our first choice for $\kmax$ is therefore given by $\kmax=\pi/R$ with $R=[3M_{\rm min}/(4\pi\bar{\rho}_m)]^{1/3}$. The results correspond to the red region bounded by continuous curves in Fig.~\ref{fig:zeta}. However, assuming an NFW profile in the HM evaluation for all halo and subhalos down to $M_{\rm min}$ implies that structures are present at even smaller scales. We will therefore consider a second definition for $\kmax$, related instead to the {\em virial} radius of the collapsed halo, or, more precisely, to $r_{200}$, corresponding to the size of spherical overdensities characterized by a mean density equal to 200 times the critical density, so that
\be\label{eq:kmaxvir}
\kmax=\left(\frac{3\,M_{\rm min}}{4\,\pi\,200\rho_{cr}}\right)^3\,.
\ee
In this case the results are shown by the red regions bounded by dashed curves. We notice that the difference between the two choices is only a factor of a few for the upper bound. However, for a closer match to the HM approach, we should in principle assume even larger values for $\kmax$ since the NFW profile describes the halo radial density in terms of a scale radius $r_s=r_{200} (M)/c_{200} (M,z)$ with the concentration parameter $c_{200}$ typically much larger than one. The redshift-dependent choice $\kmax= \pi/r_s$, with $r_s$ calculated using the recent derivation of the concentration parameter from \citet{SanchezCondePrada2013}, leads indeed to a better agreement with the HM calculation, as shown by the dotted curves in the left panel of Fig.~\ref{fig:zeta}, where the conservative uncertainty estimates of equations (\ref{eq:extrap1}) and (\ref{eq:extrap2}) are assumed. The right panels, shows instead the tighter bounds of equations (\ref{eq:extrap1}) and (\ref{eq:extrap2}), which appear to be compatible with the HM calculations in \citet{ZavalaSpringelBoylan-Kolchin2010} and \citet{AbdoEtal2010B} only for $\kmax= \pi/r_s$.

As mentioned in the previous section, however, we believe that the assumption of an NFW profile all the way down to the limiting mass can be too strong and a more effective control on the physical scales included in the HM calculation is probably required. 

We attribute to these issues the only marginal compatibility of the two methods at small redshift when the more straightforward choice of $\kmax=\pi/R$, with $R=[3M_{\rm min}/(4\pi\bar{\rho}_m)]^{1/3}$ is made. When equation~(\ref{eq:kmaxvir}) is assumed instead, the lower bound from the HM method, which is very close to the prediction of semi-analytical modeling of \citet{UllioEtal2002} shown in \citet{AbdoEtal2010B}, is well within our limits. The choice $\kmax= \pi/r_s$ leads to even closer results.

We stressed that we obtained our results without discussing {\it directly} any uncertainty on auxiliary variables such as concentration, inner halo profile, mass function, substructure, etc. A proper comparison with the uncertainty estimated with traditional methods in configuration space is limited by the choice of a {\em mass} cut-off in the HM calculation, while an extrapolation to zero distances is implicit in the assumption of an NFW profile for all halos. It is worth noting that a traditional discussion of the error budget in the HM would require much more extreme analytical extrapolations and of several different functions, not consistently defined in the literature. A simple and straightforward way to compare directly the two methods would be given by predicting the nonlinear matter power spectrum in the HM approach. Such prediction should reproduce the $P(k)$ measured in N-body simulations and provide reasonable values in the extrapolated region, thus providing an additional, important test for the method itself.

\subsection{Baryonic effects} 
\label{ssec:baryoneff}

Baryons do not trace the DM distributions through the cosmological evolution as they are influenced also by non-gravitationally interactions such as radiation losses, pressure waves and various explosive phenomena. This will obviously impact the baryonic phase-space distribution itself, but due to gravitational feedback it will also influence indirectly  the DM distribution. When including non-gravitational interactions, the close to scale invariant universality of dark matter density profiles is no longer guaranteed. Therefore there is no immediate prescription to extrapolate results from one scale to another, as for instance from cluster to dwarf galaxy scales. 

In contrast to dark matter, baryons can dissipate energy through radiation, cool and fall into the center of their surrounding dark halo. This would deepen the central  gravitational potential well, pinch the inner parts of the DM halo and hence increase the DM density. Under the assumption of circular orbits, \citet{BlumenthalEtal1986} parametrize this effect in form of adiabatic invariants. \citet{GnedinEtal2004} suggested a modified adiabatic contraction model\footnote{Other proposed analytic models are e.g. \citet{SellwoodMcGaugh2005, CardoneSereno2005, KlarMucket2008}.} to try to include the effect of non spherical orbits and better fit their hydrodynamic simulations of groups/clusters at redshift  $z \sim0$ and galaxies at redshift $z \sim 4$. Although the latter model showed better agreement with N-body simulations, other studies also found significant deviation from simulation results at galaxy scales at redshift $z=0$ \citep[e.g.][]{GustafssonFairbairnSommerLarsen2006, DuffyEtal2010}.

The presence of baryons may also act in the opposite direction, to flatten a central DM density cusp. For example, the baryons concentrated in subhalos will act as a gravitational ``glue'' to make halos resist more under tidal forces \citep{MaccioEtal2006, WeinbergEtal2008}. The resulting more efficiently transfer of angular momentum leads to dragging of DM streams out of the central parts of the main halo by dynamical friction \citep[e.g.][]{ElZantShlosmanIHoffman2001, ElZantEtal2004, ToniniLapiSalucci2006, DebattistaEtal2008, RomanoDiazEtal2008, ColeDehnenWilkinson2011}. At the same time, subhalos can suffer from increased destructive friction with the baryons in the main halo, leaving the final efficiency of DM depletion uncertain\footnote{See, for instance, the comparison of simulation results by Y. Hoffman at \href{http://chalonge.obspm.fr/CIAS10_Hoffman.pdf}{http://chalonge.obspm.fr/CIAS10\_Hoffman.pdf}.} Furthermore, stellar bar--DM interaction \citep{WeinbergKatz2002, HolleyBockelmannWeinbergKatz2005, MachadoAthanassoula2010, Sellwood2003, McMillanDehnen2005, DubinskiBerentzenShlosman2009}, and the baryon energy feedback by explosive astrophysical phenomena \citep{MashchenkoCouchmanWadsley2006, PeiraniKaySilk2008}, especially at small scales \citep{GovernatoEtal2010, PontzenGovernato2012, OhEtal2011} but also on cluster scales \citep{TeyssierEtal2011}, might significantly alter the DM distribution.

\begin{figure*}
{\includegraphics[width=0.9\textwidth]{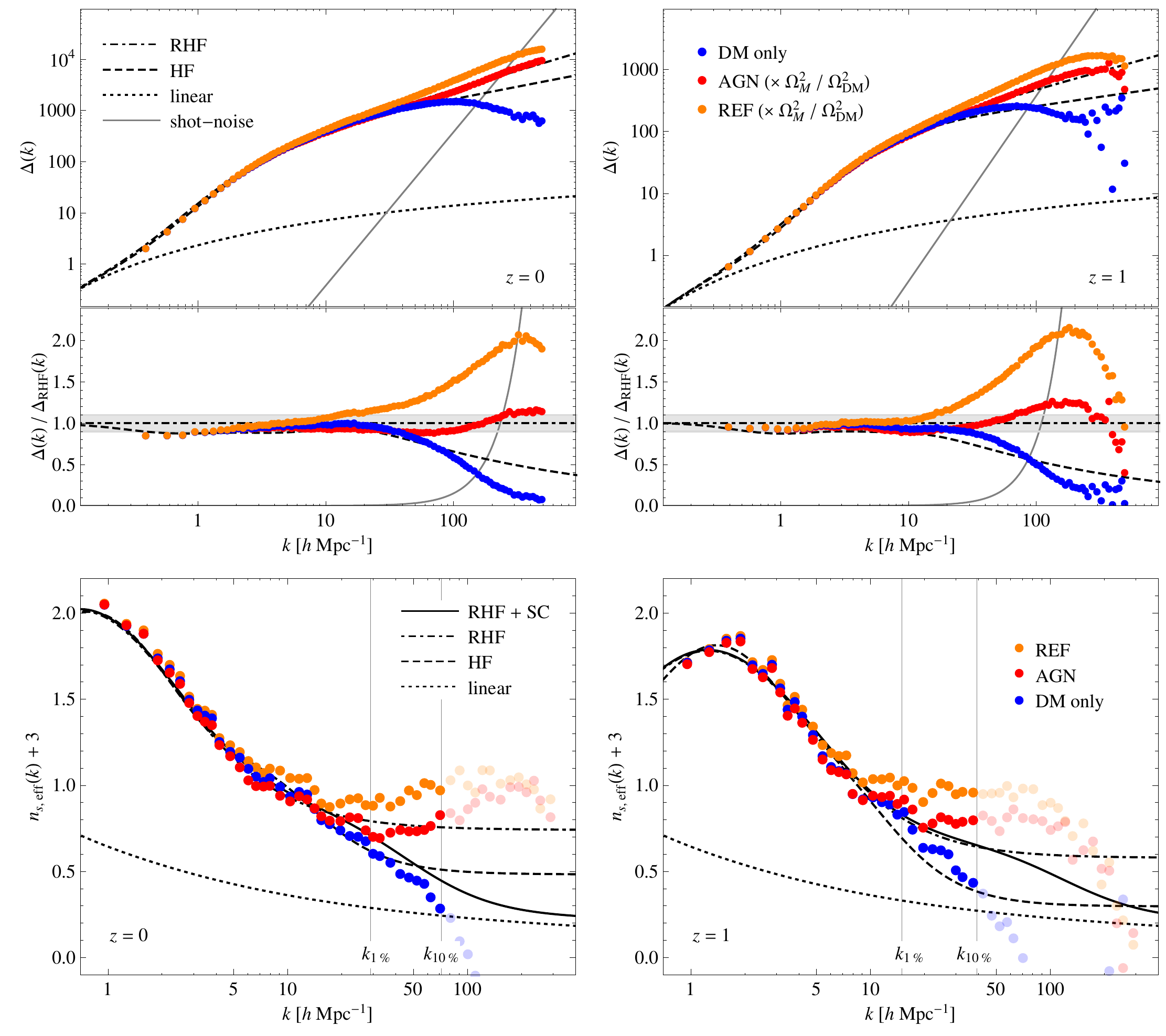}}
\caption{{\em Upper panels:} nonlinear cold dark matter power spectra from the OWLS simulations, corresponding to a pure CDM simulation (``DM only'', {\em blue dots}), a reference simulation (``REF'', {\em orange dots}) and the simulation including AGN feedback (``AGN'', {\em red dots}), see text for explanation. Numerical results are compared with the HF ({\em black dot-dashed curve}) and RHF ({\em black, continuous curve}) predictions, in addition to the linear power spectrum ({\em black, dashed}) and the shot-noise contribution ({\em black, dotted}). {\em Lower panels:} nonlinear, effective spectral index, equation (\ref{eq:ns}), obtained from the power spectrum measurements. Also shown, as vertical lines, are the scales corresponding to a 1\% and 10\% shot-noise contribution to the power spectrum of the DM only simulation. All results are shown at redshift $z=0$ ({\em left panels}) and 1 ({\em right panels}).}
\label{fig:OWLSps}
\end{figure*}

Weak lensing experiments have been an important driving force for cosmological scale N-body simulations to study the non-linear power spectrum. In order to exploit the statistical power of the future weak lensing data sets, simulators aim at modeling the power spectrum with percent accuracy, therefore naturally including also baryons in their simulations. The simulations consider a number of processes, such as radiative cooling, star formation, supernovae and feed-back from active galactic nuclei (AGN). In this work we consider the results provided by the OverWhelmingly Large Simulations (OWLS) project, which includes over 50 cosmological hydrodynamic simulations \citep{SemboloniEtal2011B}. Other simulations of similar resolution \citep[e.g.][]{RuddZentnerKravtsov2008, CasariniEtal2011B, GuilletTeyssierColombi2010} and the simulation comparison project  \citep{ScannapiecoEtal2012}) are typically consistent in their findings with the OWLS \citep[but see also recent results presented in][]{GovernatoEtal2012}.

The upper panels of Fig.~\ref{fig:OWLSps} show the nonlinear cold dark matter power spectra from the OWLS simulations. The figure is similar to Fig.~2 of \citet{VanDaalenEtal2011} which, however, shows the {\em total} matter power spectra instead, i.e. normalized to the  sum of dark matter plus baryon density. Different data points correspond to a pure CDM simulation (``DM only''), the reference simulation (``REF'') including radiative cooling and heating, and the simulation including, in addition, AGN feedback (``AGN''). See \cite{VanDaalenEtal2011} for further details. Since all simulations share the same value for the matter density parameter $\Omega_{\rm M}=\Omega_{\rm B} +\Omega_{\rm DM}$, we rescale the results of the hydrodynamical simulations (REF and AGN) by the ratio $\Omega_{\rm M}/\Omega_{\rm DM}$ to allow for a more direct comparison. Numerical results are compared with the HF and RHF predictions, in addition to the linear power spectrum and the shot-noise contribution. The lower panels show the nonlinear, effective spectral index, equation (\ref{eq:ns}), derived from the power spectrum measurements. All results are shown at redshift $z=0$ and 1. 

It is clear that the effect of radiative cooling that enables baryons to cluster on smaller scales with respect to the DM-only simulations, has, in turn, an impact on the dark matter distribution itself in the hydrodynamical simulations, providing a significant increase in power on the smallest scales probed by these measurements. Such departures from the DM-only prediction are relatively small if we consider only scales $k<k_{1\%}$\footnote{We define the scale corresponding to a given level of shot-noise contribution always for the DM-only simulation alone, in order to directly compare with the AGN and REF results.}. However, as mentioned in the introduction to this section, the effect of baryonic feedback is intrinsically scale dependent. For that reason extrapolations below the resolution of the numerical simulations are not reliable. Probably, the only robust conclusion is that the {\em minimal} effect of baryonic physics on the enhancement factor $\zeta(z)$ can be quantified at the level of 25\% for the AGN and 85\% increase for the REF run, at $z=0$, when the integral extends up to $\kmax=150\kMpc$ (a scale corresponding to a 100\% level of shot-noise contribution to the power spectrum in the DM-only case). 

In the case in which halos below the resolutions of these simulations (roughly Milky Way scale masses) are not altered much more by their baryonic component, the error calculated above might be a good proxy of the overall baryonic impact on the flux multiplier. Note that unless the profiles are extremely steep, the inner parts of the halos contribute a sub-leading amount to the full signal from the halo. For instance, it was shown in \cite{UllioEtal2002} that the cosmological signal due to dark matter annihilation changes by $\sim 20\%$ for a wide range of halo masses when the halo density profile changes from NFW to the cored Burkert \cite{Bukert1995} density profile \citep[see Fig. 3 in][]{UllioEtal2002}; so, even large modifications of the inner halo distribution of DM may not alter in a major way the expected signal. Yet, it is possible to envisage physical scenarios in which DM profiles become formally divergent at the smallest scales due to baryonic effect, for example in the case of an adiabatic contraction around intermediate mass black holes \citep{BringmannScottAkrami2012}. In such extreme scenarios the effect of ``baryons'' on the non-linear power spectrum could be significant. Even in those cases, the flux multiplier computed in this article would constitute a lower limit to the actual value.

Perhaps more worrisome is the fact that the $\Lambda$CDM shows a good agreement with data at sufficiently large scales, while discrepancies arise for fairly small galaxies, of the size of the typical dwarf galaxies of the Milky Way. It is still unclear the extent to which baryonic effects may be resolve these issues, although it is likely that they {\it can} play a role: For example,  velocity dispersions in small halos are close to 30 km s$^{-1}$, a value at which heating of the intergalactic gas by the ultraviolet photo-ionizing background should suppress gas accretion onto halos, causing them to remain dark. In addition, supernovae and stellar winds from the first generation of stars could drive remaining gas out of the shallow potential wells of these low mass halos. In most proposed explanations, the discrepancy between  $\Lambda$CDM and observations is attributed to observational bias: the underlying predictions for the dark matter clustering would be correct, but not reflected in the the properties of the ``luminous objects'' which are the observed ones. If this is the case, one can argue that baryonic effects on the DM density profile---or equivalently the concentration of these small halos---should be negligible. 

All in all, we believe that just like for the Galactic halo DM signal, also for the EDMF the main systematic uncertainty in the calculation is probably linked to baryonic effects at small scales  (kpc or below). We are confident however that the importance of the small-scale tests of cosmological models will motivate further efforts in this direction in the coming decades.

\subsection{Massive neutrinos}
\label{ssec:neutrinos}

Massive neutrinos decouple in the early Universe as relativistic particles. At a relatively large redshift---which depends on neutrino mass---they become non-relativistic, effectively contributing to the total matter component, while still having large thermal velocities. The growth of perturbations in the massive neutrinos distribution is therefore characterized by a free-streaming length corresponding to the wavenumber \citep[see, e.g.][]{LesgourguesPastor2006}
\be
k_{\rm fs}=0.018\,\Omega_{\rm M}^{1/2}\frac{m_\nu}{1\,{\rm eV}}\,\kMpc\,,
\ee
$m_\nu$ is the sum of neutrino masses, and is the main quantity affecting cosmological observables. For $k<k_{\rm fs}$, therefore in the linear regime, neutrino density fluctuations grow as the cold dark matter ones. At $k\gg k_{\rm fs}$ free-streaming dumps neutrino perturbations inducing, in turn, a back-reaction on cold dark matter described in linear theory quantifiable as
\be
\frac{P_{{\rm DM}, m_\nu}(k)}{P_{{\rm DM}, m_\nu=0}(k)}\simeq 1-6 \frac{\Omega_\nu}{\Omega_{\rm M}}\,,
\ee
$\Omega_\nu\sim m_\nu / (93\,h^2 {\rm eV})$ being the massive neutrino contribution to the total energy density. The nonlinear evolution of the matter power spectrum has been investigated most recently with numerical simulations in \citet{BirdVielHaehnelt2012} and \citet{WagnerVerdeJimenez2012}, where massive neutrinos are described as particles but with an initial velocity component drawn from the appropriate Fermi-Dirac distribution. These authors focused their attention on the total matter power spectrum, rather than the cold dark matter one, noticing that the suppression of power at scales of the order of $k\sim 1\kMpc$ for $z=0$ is larger than linear theory predicts. At slightly smaller scales, but still $k\le 10 \kMpc$, the suppression approaches again the one predicted by linear theory \citep[see Fig.~2 in][]{BirdVielHaehnelt2012}~\footnote{Note that the authors also propose a modified version of the \texttt{halofit} code that accurately describes these features. This is achieved, however, by modifying as well the asymptotic behavior of the formula for $k\rightarrow 0$ and we cannot therefore expect an extrapolation beyond $k\sim 10\kMpc$ to be of particular significance. In addition the \texttt{halofit} predictions for the cold dark matter component alone, have not been properly tested.}. The overall effect is typically of the order of 10--30\%, for allowed values of the total neutrino mass \citep[see e.g.][for current constraints]{Planck2013parameters}, which we believe is a good proxy for the maximal expected uncertainty on the flux multiplier due to neutrinos. Needless to say, numerical simulation of greater resolution would be helpful in exploring the deeply nonlinear regime, but one can argue that the effects of massive neutrinos on the flux multiplier should be significantly less important than baryonic ones. Due the large thermal velocities characterizing their distribution in phase space, only a small fraction of the neutrino number participates in the virialization of structures \citep{Villaescusa-NavarroEtal2013}. The impact of both linear and nonlinear perturbations on the cold dark matter distribution at very small scales are expected to be negligible, compared to the order-of-magnitude uncertainties coming from the effects discussed in Sec.~\ref{ssec:resolution} and Sec.~\ref{ssec:baryoneff}.

\subsection{Upper limit on the integrated power spectrum from supernovae lensing magnification}
\label{ssec:SN}

The daring extrapolation required by the evaluation of the flux multiplier $\zeta(z)$ is particularly significant given the fact that we have little information on the highly nonlinear regime of the matter power spectrum, even indirectly, from observations. Even a relatively loose upper limit to the amplitude of fluctuations at very small scale could limit the large uncertainty crudely estimated in the previous section. 

One such upper limit can be obtained from the measured scatter in the magnitude of standard candles as Type Ia supernovae. This quantity, which is a direct observable, is the result of several systematic and statistical effects whose specific nature is irrelevant here. We notice, in fact, that one particular source of scatter is the magnification of the supernovae luminosity due to weak lensing by the matter distribution along the line-of-sight. The distance modulus $\mu(z)$ of a SNIa at redshift $z$ can be written as \citep[see, e.g.][]{DodelsonVallinotto2006, MunshiValageas2006}
\be
\mu(z)=\mu_0(z)+C+\delta\mu_{\rm int}+\delta\mu_{\rm WL}\,,
\ee
where $\mu_0$ is the unlensed distance modulus related to the luminosity distance $d_L$ as 
\be
\mu_0(z)=5\ln 10\left[\frac{d_L(z)}{10\,{\rm pc}}\right]\,,
\ee
$C$ is a constant related to mean absorption and calibration, while $\delta\mu_{int}$ quantifies intrinsic fluctuations, including but not limited to luminosity dispersion. In addition, the term $\delta\mu_{\rm WL}$ accounts for weak lensing effects. While the mean is given by $\langle\mu(z)\rangle=\mu_0(z)+C$, the variance of the observed $\mu(z)$ includes the contribution from weak lensing, that is
\bea
\sigma_{\rm WL}^2(z)\!\!&\!\! =\!\! &\!\! \frac{225\,\pi\,\Omega_{\rm M}^2\,H_0^4}{4\,(\ln 10)^2}\,\int_0^{\chi(z)}\!\!\!\!d\chi'\,[1+z(\chi')]^2\frac{\chi'^2\,(\chi-\chi')^2}{\chi^2}\nonumber\\
& & \times\int_0^\infty dk\,\frac{\Delta(k,\chi')}{k^2}\,,
\eea
where $\chi(z)$ is the comoving distance along the line-of-sight to redshift $z$. This expression amounts essentially to a weighted integral over the matter power spectrum $\Delta(k)$. Given the observed variance $\sigma_{\rm obs}$, we can place an upper limit such integral simply by imposing $\sigma_{\rm WL}\le\sigma_{\rm obs}$. 

Here we want to check whether such limit does translate to an interesting bound on $\Delta(k)$, particularly since the integral above extends to all values of $k$, although down-weighting the contribution at large $k$ by the $1/k^2$ factor. Note that by \citet{BenDayanKalaydzhyan2013} its was recently argued that similar considerations can already place tight constraints (comparable or better than PLANCK ones) on cosmological parameters such as the ``running of the running'' of the spectral index.

\begin{figure*}
{\includegraphics[width=0.9\textwidth]{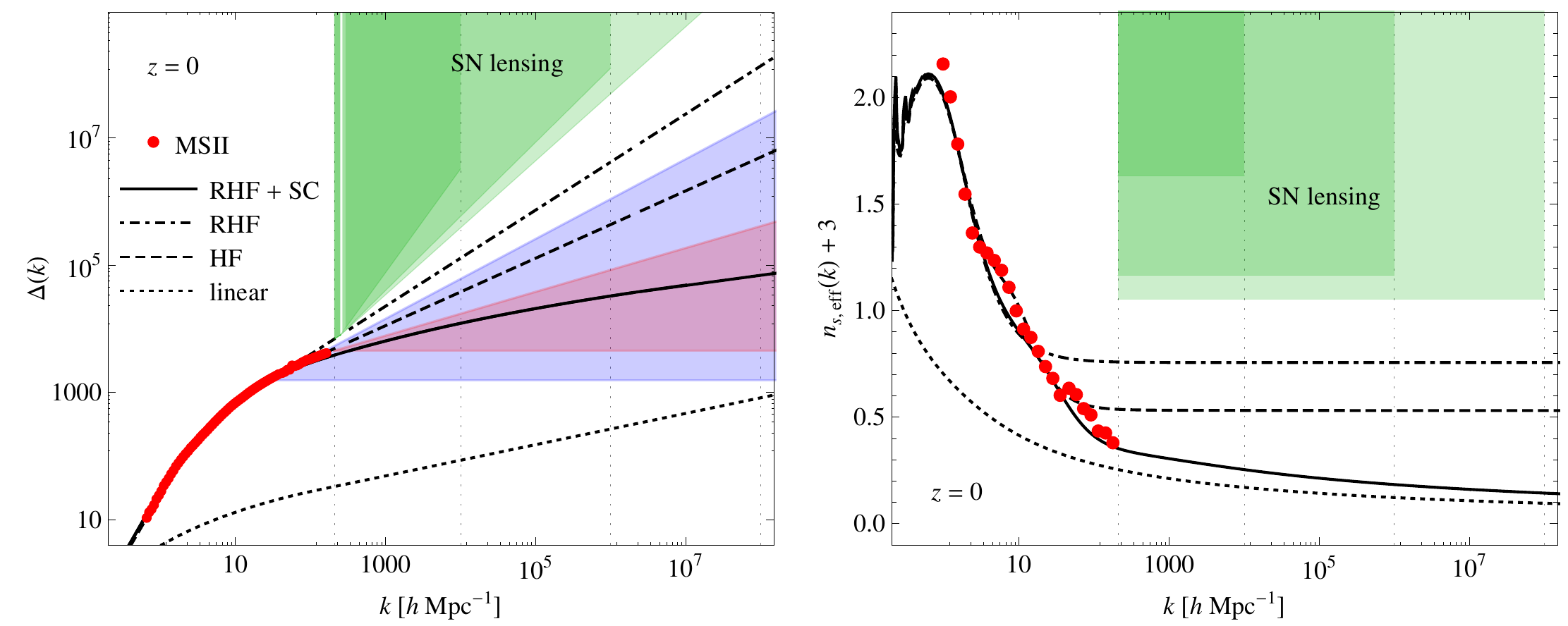}} 
\caption{Comparison between the MSII data and extrapolations for power spectrum ({\em left panel}) and effective  spectral index ({\em right panel}) with the regions probed by SN scatter from weak lensing magnification. Such regions assume supernovae at redshift $z=1$ and $k_{\rm max}=10^4$, $10^6$ and $10^8\kMpc$ ({\em darker to lighter green areas, respectively}). \label{fig:SN}}
\end{figure*}

Since we aim at a rough assessment, we assume a toy model for the power spectrum $\Delta(k,z)$ given by the RHF formula up to a scale $k_\star=k_{1\%}$ where the shot-noise contribution is the one of the MSII simulation, while for $k>k_\star$ we consider the simple power-law
\be
\Delta(k)=\Delta_{\rm RHF}(k_\star)\left(\frac{k}{k_\star}\right)^\alpha\,,
\ee
with $\alpha$ a free constant. We assume, in other terms, a constant value $\neff=\alpha-3$ for the effective spectral index beyond $k_\star$. The parameter $\alpha$ will be constrained by the relation $\sigma_{\rm WL}\le\sigma_{\rm obs}$, where we assume $\sigma_{\rm obs}=0.1$ \citep[see, e.g.][]{SuzukiEtal2012} at a fixed redshift of $z=1$. For the integration we consider three different values of $k_{\rm max}$ given by $10^4$, $10^6$ and $10^8\kMpc$. The results are shown in Fig.~\ref{fig:SN}, compared to the extrapolations of HF, RHF and RHF corrected by stable clustering, both in terms of $\Delta(k)$ as in terms of the effective spectral index $\neff(k)$. 

The regions probed by the scatter in supernovae magnitudes are several order of magnitudes above expectations for $k_{\rm max}=10^6\kMpc$ and are therefore not providing any significant constrain to our uncertainty estimate (note that the same region is shown as well on the right panel of  Fig.~\ref{fig:zeta}). Note however that these regions are only a factor of a few larger than the largest Halo Model predictions of \cite{ZavalaSpringelBoylan-Kolchin2010}.  Further, more rigorous investigations of this interesting possibility to provide an upper bound on the flux multiplier $\zeta(z)$ might be considered in the future, particularly in the event that improved knowledge on baryonic physics would suggest substantial changes to the simple picture provided by the stable clustering assumption in the highly nonlinear regime.

\section{Other cosmological uncertainties}
\label{sec:cosmouncertainty}

Although the small-scale behavior of the power spectrum and the impact of baryons at those scales certainly provide the major uncertainty in the estimate of the flux multiplier, additional uncertainties exist that affect the EDMF, both via $\zeta(z)$ and via other factors. In the following, we analyze the two main effects neglected in the previous discussion, notably: the statistical uncertainties in cosmological parameters (Section~\ref{ssec:statA}), and the accuracy with which the EBL is empirically known as well as theoretically modeled to determine optical depth parameter $\tau$ (Section~\ref{ssec:statB}).

Also, while we concentrate here on the annihilating dark matter case, it is worth noting that the uncertainties in $e^{-\tau}$ and the $\psi(z)$ prefactor defined in equation~(\ref{psiz}) below are the only uncertainties entering the average flux signal for {\it decaying} DM (see for example equation~(5) in \cite{CirelliPanciSerpico2010}), lacking the analogous of the $\zeta(z)$ term. This is the formal reason why the computation of the flux from decaying DM is much more robust.

\subsection{Statistical uncertainties in cosmological parameters}
\label{ssec:statA}

Not surprisingly, statistical uncertainties in the cosmological parameters, either in the $\Lambda$CDM model or in extensions of it, can moderately affect the computation of $\zeta(z)$. To estimate these effects, we evaluated the logarithmic derivatives:
\be
\varphi_i (z,{\bf y}_0) \equiv  \frac{\partial \ln \zeta(z,{\bf y})}{\partial \ln y_i}\Bigg |_{\bf y=y_0}\,.
\ee
where ${\bf y_0}$ indicate the cosmological parameters at their best fit value. By multiplying $\varphi_i$ times the (current) fractional error on the cosmological parameter $y_i$ (which we take from 68\% CL uncertainties in Planck results \citep{Planck2013parameters}) one can quickly estimate the relative error induced by that uncertainty at the linear level.   

We evaluated the above quantities $\varphi_i$  at $z=0$ by fixing the value of the cosmological parameters at their fiducial values ${\bf y}_0$ as assumed in the Millennium Simulation. In particular, within flat $\Lambda$CDM model we considered the baryon density ($\Omega_{\rm B}$), the dark matter density ($\Omega_{\rm DM}$),  the reduced (dimensionless) Hubble rate $h$, the normalization ($\Delta^2_{\cal R}$) and the spectral index of the primordial power spectrum ($n_s $), as well as the running of the spectral index ($d n_s/d \log k $) and the dark energy equation of state parameter $w$ in extended cosmological models. The latter two parameters also give an idea of the ``theoretical cosmological model'' error in the estimate of these statistical uncertainties.
 Finally, we considered two values of $k_{\rm max}=100\kMpc$ and $k_{\rm max}=10^6\kMpc$, to gauge if the dependency from some parameters is crucially altered by the power-spectrum extrapolation.

We summarize our main results below:
\begin{itemize}
\item while there is a non-negligible sensitivity of the signal for most parameters of the minimal concordance cosmology, this sensitivity does not necessarily translate into major uncertainty in the signal, because the corresponding parameters are currently known with good precision. Errors are well below 10\%.
\item the effect of the uncertainty on the parameter $n_s$ significantly grows with $k_{\rm max}$. This is expected, since it controls the ultraviolet tail of the power spectrum. Uncertainties up to the $\sim$10\% level are reasonable for $k_{\rm max}\sim 10^6\kMpc$.
\item moving beyond the minimal concordance cosmology,  two effects arise: on one hand, the uncertainties on the determination  (e.g. from fit to CMB data) of basic cosmology parameters grow.
Second and more important, other cosmological parameters of the extended models can have a significant impact: for example, in  wCDM cosmologies, the induced error on the flux multiplier is comparable to the relative error on $w$ (currently at the 10\% level). This remark applies in particular to cosmologies presenting a modification of the primordial power spectrum. For example, for realistic large values of  $k_{\rm max}$, the relative error on $\zeta$ is comparable to the one on the spectral running \citep[currently almost at the 50\% level at 1-$\sigma$,][]{Planck2013parameters}.
\end{itemize}
The impact on $\zeta(z)$ is not the only one: from equation~\ref{finaleq} it is clear that the EDMF depends from cosmology also via the pre-factor
\be
\psi(z)\equiv  \frac{\Omega_{\rm DM}^2}{H(z)}\,.\label{psiz}
\ee 
The dependence of the EDMF on the Hubble rate normalization $h$ via this $\psi(z)$ is trivial (inverse proportionality). We can neglect here the few percent-level uncertainty in the ($z-$independent) normalization due to this  parameter.  On the other hand, even in $\Lambda$CDM, the dependence on $\Omega_{\rm DM}$ is larger: $d \ln \psi/d \ln \Omega_{\rm DM}\simeq 1.5\div 2$. For current uncertainty in $\Omega_{\rm DM}$, this translates into a $\sim 10\%$ uncertainty in $\psi$. Note also that, although this error slightly depends on $z$, this dependency is about five times smaller than the normalization uncertainty, and its impact on the energy dependence of the EDMF can be safely neglected.

In summary, statistical uncertainties in the parameters of the $\Lambda$CDM model have only a small impact on the EDMF, at the  ${\cal O}(10\%)$ level and often smaller. However, moving away from the minimal concordance model (e.g. allowing for a spectral running) can bring the statistical uncertainty associated to cosmological parameters to larger values, up to  a factor $\cal{O}$(2). Nonetheless, like for the case extrapolation of the power spectrum to large$-k$, most of this effect concerns the {\it normalization} of the signal. Given that the extrapolation errors covered in the previous sections lead to at least one order of magnitude uncertainty, these statistical errors are clearly sub-leading and will not be considered in further detail.

\subsection{EBL-related uncertainty}
\label{ssec:statB}

The other important factor entering the formula of  equation~(\ref{finaleq}) is the $e^{-\tau(E,z)}$ accounting for absorption of high energy photons onto the extragalactic background light. This is more worrisome since it clearly concerns the  {\it spectral shape} of the EDMF, which is only affected to a lesser extent by the effects we have previously treated. 

\begin{figure*}
{\includegraphics[width=0.9\textwidth]{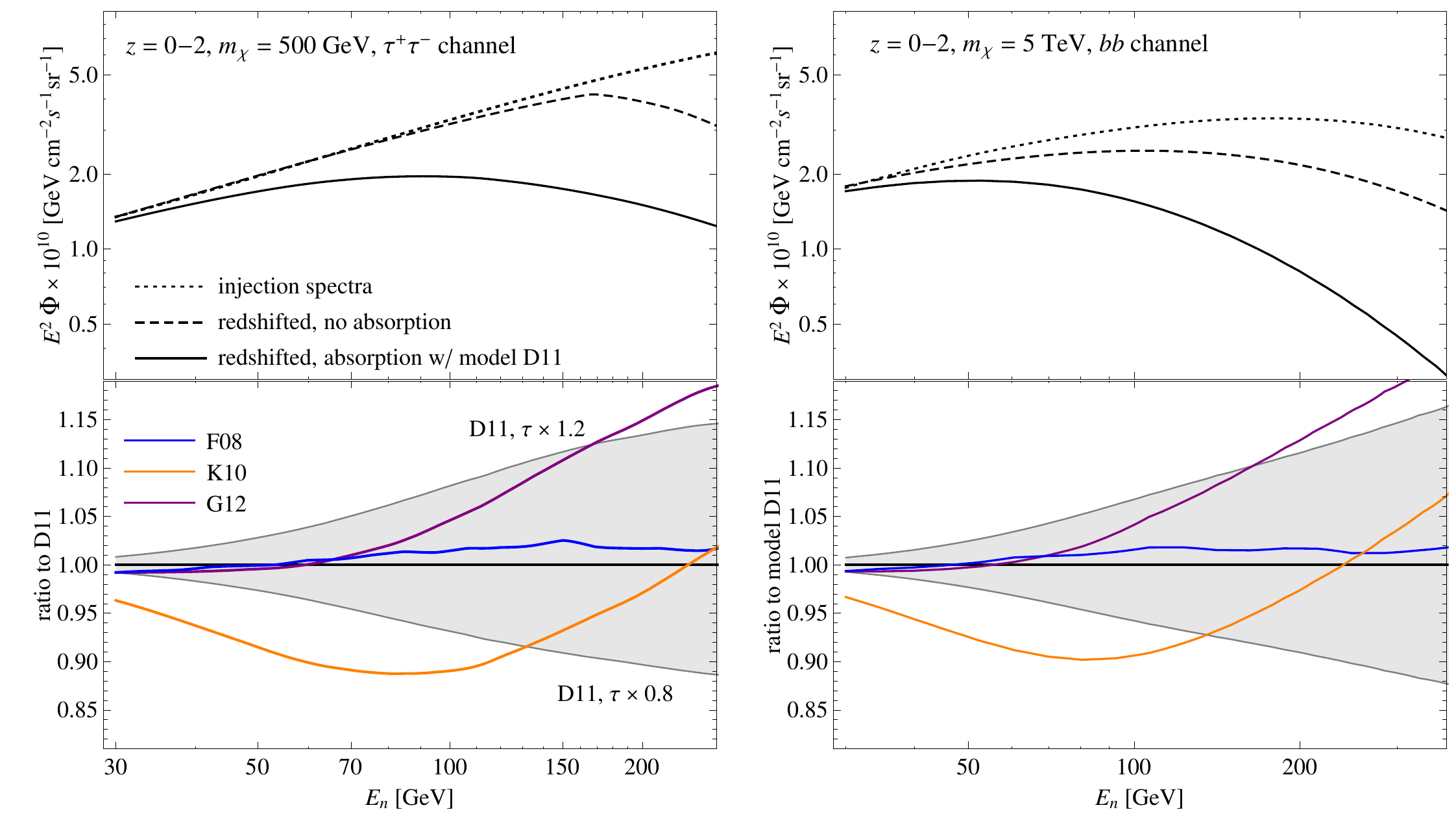}}
\caption{{\em Upper panels}: injection spectrum from DM annihilation calculated with the \texttt{PPPC4DMID} code of \citet{CirelliEtal2011} ({\em dotted curve}), the same spectrum as observed on the Earth if originating from the cosmological DM annihilations, i.e. including redshift effects ({\em dashed curve}) and the spectrum at Earth when also the EBL absorption within the model D11 \citep{DominguezEtal2011} is accounted for ({\em solid curve}). {\em Lower panels}: Ratio of the Cosmological DM annihilation spectrum including EBL calculated in different models F08 \citep{FranceschiniRodighieroVaccari2008}, k10 \citep{KneiskeDole2010} and G12 \citep{GilmoreEtal2012} to the D11 case. The shaded area is bounded by the D11 prediction when the function $\tau(E,z)$ is enhanced or lowered by $20\%$. \label{fig:EBLratio}}
\end{figure*}

Fortunately, over the last few years the knowledge of the EBL background has greatly improved, in particular thanks to gamma-ray observations. Notably, both the Fermi-LAT collaboration \citep{AckermannEtal2012} and the H.E.S.S. collaboration \citep{AbramowskiEtal2013} have reported a firm detection (above 5 $\sigma$) of the EBL absorption effect. Essentially, the absorption effect has been measured at a level which validates the ``modern'' predictions, starting from \citet{FranceschiniRodighieroVaccari2008} (F08) till the more recent models of \citet{DominguezEtal2011} (D11) and \citet{GilmoreEtal2012} (G12) or the minimal model of \citet{KneiskeDole2010} (K10). For what concerns Fermi, the best fit values are close to fiducial predictions, with  1 $\sigma$ error  of about 20\%;  the H.E.S.S. ``high energy'' analysis finds  best fit values only slightly larger than the one of~\cite{FranceschiniRodighieroVaccari2008} and consistent with that within the errors,  which amount to about 30\% at 1 $\sigma$.

In order to quantify the spectral uncertainty effect due to limited knowledge of the EBL,  we present some examples of the spread in energy shape due to different EBL absorption models for representative benchmark  spectra. In particular we consider the spectra of a 500 GeV particle annihilating  into the $\tau^+ \tau^-$ channel and a 5 TeV particle annihilating  into the $b{\bar b}$ channel. We focus on a high DM mass range where the effect is pronounced, as absorption is more relevant for gamma-ray energies roughly above 50 GeV. We also note that while the current measurement of the isotropic gamma-ray background by the Fermi-LAT extends up to 100 GeV \citep{AbdoEtal2010A} and is only weakly sensitive to $\gsi$TeV DM candidates, those masses are within the reach of the new measurement (extending up to 820 GeV) preliminarly presented by the \citet{AckermannEtal2014}. As our benchmark absorption model we take the one of D11.

In the upper panels of Fig.~\ref{fig:EBLratio} we show the injection spectra for our benchmark DM models, and the resulting cosmological DM spectra, modified due to the effects of cosmological redshift and absorption. In the lower panels we show the ratio of the cosmological spectra with EBL absorption calculated in different models mentioned above to our benchmark EBL model. We do so in the $z$ and $E$ range for which the EBL data for all of the models are available (i.e. $z=0\--2$, $E\gsi 30$ GeV). In addition, within the D11 absorption model we increase (and decrease) the function $\tau(E,z)$ by 20\% independently of the energy and redshift, to mimic the {\it experimental} uncertainty on this parameter. We find that the overall uncertainty is typically within 10\% and not greater than 20\% for the redshift and energy range of interest. Note also that the uncertainties are modest (${\cal O}$(5\%)) at energies of tens of GeV, and only grow to the 10-20\% level at hundreds of GeV, where one probes anyway the decreasing tail of the spectrum, plausibly affected by larger experimental errors (e.g. reduced statistics and worsened rejection capability). Hence, we conclude that---at least for current experimental uncertainties---the limited knowledge of the EBL only marginally degrades the precision with which the spectrum of the expected EDMF can be computed.  

A final comment is in order:  the energy initially stored in the absorbed photons also eventually contributes to the gamma-ray signal: the {\cal O}(10-100) GeV secondary $e^\pm$ pairs lose a (usually large) fraction of their energy via inverse-compton scattering of CMB photons, up-scattering producing ${\cal O}$(0.1-1) GeV tertiary photons. We neglect it here since, to a large extent, it does not produce independent uncertainties: neglecting all other sub-leading energy losses (synchrotron losses, gas heating), this is an exactly computable contribution that only depends on $\zeta(z)$ and $\tau(z)$.

\section{Summary and conclusions}
\label{sec:conclusions}

In this article we have presented new  computations and error estimates of the so-called flux multiplier, the crucial ingredient for the estimate of the Extragalactic DM annihilation Flux (EDMF). This signal is usually calculated in the Halo Model (HM) framework, but as we already argued in~\citet{SerpicoEtal2012}, a more direct evaluation is possible solely based on the (non-linear) DM power spectrum. This followup paper uses that approach to provide a critical overview of the different steps and assumptions entering the computation of the EDMF, with a particular emphasis on the assessment of the uncertainties involved. 

In the context of  cosmologies where DM constitutes the only matter component, the major results of the paper are reported in Sec. IIIA and can be summarized as follows:
\begin{itemize}
\item Based on the {\it linear} power spectrum, the modified \texttt{halofit} fitting function provided in~\cite{TakahashiEtal2012} and the principle of stable clustering, we provide a best guess estimate for the full non-linear power spectrum, equation~(\ref{eq:RHFSC}), down to small scales. 
\item By comparing two consistent cosmological simulations at different resolution, Millennium Simulation (MS) and Millennium Simulation II (MSII), we provided a more solid criterion to estimate a realistic uncertainty due to extrapolation of simulations to yet unresolved very small scales, see equations~(\ref{eq:extrap1}) and (\ref{eq:extrap2}), the less conservative assumptions of equations~(\ref{eq:extrap1C}) and (\ref{eq:extrap2C}), and the discussion about the best guess for $k_\star$.
\item A rough comparison showed that our results are consistent with those obtained with typical halo model predictions. Also, the estimated uncertainty is narrower at low $z$ and wider at larger $z$, under our most conservative assumptions. We stressed how a direct comparison is made difficult by some ambiguities in the mapping from a ``power spectrum'' framework  to a ``halo model'' framework especially at scales close to the cutoff ($\kmax$ vs. $M_{\rm min}$). Depending on the prescription used, the PS best guess for extra galactic DM annihilation flux is lower than typical HM predictions.
\end{itemize}

It is also worth mentioning that the strategy for the evaluation of the signal and error outlined in this article has the advantage of being easily improved when simulations of higher resolution become available.  The reduction of the estimated uncertainty in passing from MS to MSII shown here is a clear manifestation of this advantage.  Actually, already with current simulations, having a finer grid in $z$ from MSII would likely translate into more stringent estimates of $\zeta(z)$ especially at high $z$, as it is clear from  equation (\ref{eq:extrap2}) and the results shown in Fig.s~\ref{fig:extrap} and~\ref{fig:zeta}. On the other hand, HM computations of the flux multiplier face some difficulties of both conceptual and practical nature: i) They must proceed via the determination from simulations of several auxiliary quantities which are not {\it separately} entering the flux multiplier calculation. Establishing for example the correlation between errors in the concentration parameter determination and the mass function ones might not be trivial. Usually, different parameters are varied independently when estimated errors in the prediction, which may be partially responsible for why higher values of the flux multiplier can be found in HM based computations. ii) Closer to the cutoff in the power-spectrum due to the physical nature of DM, the very  notion of ``halo'', with a (quasi)universal density profile may break down. We discussed some of the difficulties in dealing with this issue in a proper way, which also prevents one to go beyond rough comparisons between the two methods. iii) HM-based computations have to face several ambiguities and deal sometimes with non consistently defined quantities, varying from one simulation group to another. It is enlightening to report here one of the many similar statements that we found in the literature~\citep{KlypinTrujilloGomezPrimack2010}:

{\em The virial mass is a well defined quantity for distinct halos (those that are not subhalos), but it is ambiguous for subhalos. It strongly depends on how a particular halo finder defines the truncation radius and removes unbound particles. It depends on the distance to the center of the host halo because of tidal striping. [\ldots] Even for distinct halos the virial mass is an inconvenient property because there are different definitions of ``virial mass''. None of them is better than the other and different research groups prefer to use their own definition. This causes confusion in the community and makes comparison of results less accurate.}

Also, we provided a first assessment of the additional errors introduced by neglecting the role of baryons and neutrinos, establishing that the former is clearly more relevant. Using the OWLS simulations, we estimated a lower-limit on the uncertainty induced by baryons of almost a factor of two, based only on the structures resolved by these simulations. We argued that the understanding of the baryonic effects at different scales provides the major {\em systematic} obstacle towards a refinement of the computation of the flux multiplier. In principle, very large enhancements of the signal can be obtained in extreme cases. In this context, however, an empirical upper limit on the magnitude of the small-scale power spectrum might be useful. We inferred one such limit from the measured variance in supernovae Ia lensing magnification. Albeit too loose to constrain most of the current predictions, refinements of these kind of constraints may provide a useful cross-check of some extreme models, in the future.

Finally, we have addressed the statistical errors on the EDMF due to cosmological parameters present in a standard cosmology (and in some minimal extensions), as well as the errors induced by the current knowledge of the extragalactic background light (which is responsible for a finite absorption probability of gamma-ray photons of extragalactic origin). Both effects appear however sub-leading, if compared with systematic uncertainties previously mentioned. 

Globally, for typical WIMP parameters current uncertainties on the EDMF mostly enter the normalization of the signal (via small scale extrapolation or baryonic effects), the spectral-shape uncertainty being relatively more modest. Uncertainties due to cosmological parameters determined from large-scale cosmology probes have a negligible effect, with the possible exception of extra parameters in non-minimal cosmologies affecting the power spectrum (notably the spectral running, among the parameters we tested). Needless to say, both errors on the normalization and the $z-$dependence of the flux multiplier should be considered {\it lower limits} to the current uncertainties. Yet, provided that other poorly understood effects do not introduce a significantly large redshift-dependence in the flux multiplier,  one can still envisage a strategy where the EDMF could be used as  independent confirmation of a gamma-signal detected elsewhere, for instance from the inner halo of the Milky Way. To that purpose, a {\it spectral} template, based on the putative DM signal, could be used to fit the high-latitude residuals with a free normalization.  Here we do not discuss these issues in any quantitative details, but limit to note that in general the extragalactic signal at high latitudes is accompanied by a comparable Galactic DM signal, see e.g. Fig.~1 in~ \citet{HooperSerpico2007}. While this may complicate the actual analysis, a synergic treatment of both signals may improve the chances of detection.

\section*{Acknowledgments}

We thank C. Giocoli, M. Viel and F. Villaescusa-Navarro for useful discussions.  We are  particularly grateful to R. Angulo, M. Boylan-Kolchin and J. Schaye for sharing with us with the MS and OWLS power spectrum measurements and to A. Taruya and T. Nishimichi for timely providing us with a helpful implementation of their revised \texttt{halofit} formula. E.S. especially thanks P. Valageas for pointing out SN observation as a possible test of the nonlinear power spectrum. M.G. is supported by the Belgian Science Policy (IAP VII/37), the IISN and the ARC project ``Beyond Einstein: fundamental aspects of gravitational interactions". D.T. is partially supported by CONACyT grant 151234, and is thankful to IFUNAM and the Abdus Salam ICTP for support and hospitality during his contribution to this work. 

\medskip

{\it Note added:} While this work was being finalized, the preprint of \citet{FedeliMoscardini2014} appeared: that article performs an analysis similar to the one presented in section Sec.~\ref{ssec:SN}, albeit from a complementary perspective, in the framework of the Halo Model.

\end{document}